\documentclass[english, aps,prd,twocolumn,twoside,superscriptaddress,floatfix, nofootinbib]{revtex4-1}

\usepackage[utf8]{inputenc}
\usepackage{amsmath}
\usepackage{geometry}
\usepackage[dvipsnames]{xcolor}
\usepackage{graphicx}
\usepackage{float}
\usepackage{lipsum, babel}
\usepackage{graphicx}
\usepackage{color}
\usepackage{hyperref}
\usepackage{ifthen,amsmath,amssymb, bm}

\newcommand{\beq} {\begin{equation}}
\newcommand{\eeq} {\end{equation}}
\newcommand{\bal} {\begin{aligned}}
\newcommand{\eal} {\end{aligned}}

\definecolor{matplotlib_blue}{RGB}{44, 160, 44} 
\definecolor{matplotlib_red}{RGB}{214, 39, 40} 
\definecolor{matplotlib_orange}{RGB}{255, 127, 14} 
\definecolor{matplotlib_purple}{RGB}{148, 103, 189} 
\definecolor{matplotlib_blue}{RGB}{31, 119, 180} 

\begin{document}

\title{Foreground-immune CMB lensing reconstruction with polarization}

\author{Noah Sailer}
\email{nsailer@berkeley.edu}
\affiliation{Berkeley Center for Cosmological Physics, Department of Physics,
University of California, Berkeley, CA 94720, USA}
\affiliation{Lawrence Berkeley National Laboratory, One Cyclotron Road, Berkeley, CA 94720, USA}
\author{Simone Ferraro} 
\email{sferraro@lbl.gov}
\affiliation{Lawrence Berkeley National Laboratory, One Cyclotron Road, Berkeley, CA 94720, USA}
\affiliation{Berkeley Center for Cosmological Physics, Department of Physics,
University of California, Berkeley, CA 94720, USA}
\author{Emmanuel Schaan}
\email{eschaan@slac.stanford.edu}
\affiliation{SLAC National Accelerator Laboratory, Menlo Park, CA 94025, USA}
\affiliation{Kavli Institute for Particle Astrophysics and Cosmology and Department of Physics, Stanford University, Stanford, CA 94305, USA}

\begin{abstract}
Extragalactic foregrounds are known to generate significant biases in temperature-based CMB lensing reconstruction. Several techniques, which include ``source hardening'' and ``shear-only estimators'' have been proposed to mitigate contamination and have been shown to be very effective at reducing foreground-induced biases. 
Here we extend both techniques to polarization, which will be an essential component of CMB lensing reconstruction for future experiments,
and investigate the ``large-lens'' limit analytically to gain insight on the origin and scaling of foreground biases, as well as the sensitivity to their profiles.
Using simulations of polarized point sources, we estimate the expected bias to both Simons Observatory and CMB-S4 like (polarization-based) lensing reconstruction, finding that biases to the former are minuscule while those to the latter are potentially non-negligible at small scales ($L\sim1000-2000$). In particular, we show that for a CMB-S4 like experiment, an optimal linear combination of point-source hardened estimators can reduce the (point-source induced) bias to the CMB lensing power spectrum by up to two orders of magnitude, at a $\sim4\%$ noise cost relative to the global minimum variance estimator.
\end{abstract}

\maketitle

\section{Introduction}
Point sources and other foregrounds induce biases \cite{2014ApJ...786...13V, 2018PhRvD..97b3512F} to standard quadratic estimators of the weak lensing potential due to the fact that foregrounds are non-Gaussian and correlated with the lensing potential. The impact of foregrounds on the CMB temperature lensing reconstruction has been explored extensively in the literature and a number of mitigation techniques have been proposed. These range from decreasing the multipole range used in the reconstruction to more sophisticated ``geometric methods'' that employ the different symmetries of foregrounds and lensing.

Hardening against point sources in temperature-based reconstruction has been proposed in \cite{2013MNRAS.431..609N,2014JCAP...03..024O,Namikawa:2013xka, Sailer:2020lal} and has been shown to reduce the foreground biases by an order of magnitude with very modest noise cost of order $10\%$. A generalization to hardening against a fixed profile (profile hardening) was proposed in \cite{Sailer:2020lal} and has been shown to be even more effective when the mean profile of tSZ halos is hardened against.
An even more general ``foreground-immune'' estimator is the shear-only estimator \cite{2019PhRvL.122r1301S}, which suppresses the response to arbitrary azimuthally symmetric profiles. Further generalizations to ``multipole'' estimators have also been proposed in \cite{2019PhRvL.122r1301S}.

Moreover, frequency-based methods to reduce biases have been proposed in \cite{2018PhRvD..98b3534M, Sailer:2021vpm}, and the optimal combination of multi-frequency and geometric methods have been explored in \cite{Darwish:2021ycf}. Extragalactic foregrounds can also affect delensing of the $B-$modes produced by primordial gravitational waves, and the techniques mentioned above appear to be very effective at reducing these biases \cite{BaleatoLizancos:2022vvr}.

In this paper we extend two of the most promising geometric techniques used in temperature reconstruction (i.e. bias hardening and shear-only reconstruction) to polarization. We note that these can be used in addition to multi-frequency foreground reduction to further mitigate foreground biases. 

CMB polarization dominates the statistical power of lensing reconstruction for map noise $\Delta_T \lesssim 5\,\mu$K-arcmin, and therefore will be the principal channel in the future. For current experiments, polarization already dominates reconstruction for SPT-3G \cite{2014SPIE.9153E..1PB} and will play an important role in Simons Observatory \cite{2019JCAP...02..056A} and CMB-S4 \cite{CMB-S4:2022ght}. It is therefore important to characterize and mitigate possible biases induced by foregrounds in polarization.

First of all, we note that extragalactic foregrounds in polarization are thought to be simpler than in temperature. The largest sources of extragalactic foregrounds, such as emission from the Cosmic Infrared Background (CIB) or the thermal Sunyaev-Zel'dovich effect (tSZ), are mostly unpolarized (although a small polarized signal is present at higher order \cite{Feng:2019miy, Deutsch:2017cja}), and therefore their effect on lensing reconstruction from polarization is likely negligible. However, bright infrared and radio point sources are known to be polarized at the $\sim$ few \% level  \cite{Li:2021ial, Datta:2018oae, SPT:2019wyt}, inducing potential biases.  Therefore we'll concentrate on mitigating the effect of polarized point sources, while noting that the techniques we present here have also been shown to be very effective against CIB and tSZ in temperature, and therefore are likely to also work more generally in polarization.

In this paper we focus on extragalactic foregrounds. The methods developed to mitigate these biases may also prove useful for galactic foregrounds, whose biases are potentially important and their study will be the subject of future work. 

The remainder of the paper is organized as follows: in section \ref{sec:theory}, we outline the theory of quadratic estimators, including multi-frequency and minimum-variance combinations. In section \ref{sec:bias_hardening}, we extend the bias hardening techniques to polarization, while in section \ref{sec:LLS} we explore generalizations to the ``shear-only'' estimators, highlighting important differences in polarization. In section \ref{sec:noise} we study the noise properties of these new estimators and in section \ref{sec:bias} we investigate the bias reduction obtained. We conclude in section \ref{sec:conclusions}. 

\section{Linear responses and quadratic estimators}
\label{sec:theory}
Here we review the basic theory of quadratic estimators in temperature and polarization. We separately consider the cases of reconstructing the convergence from a single pair of maps and the case of summing over all pairs, and derive the minimum variance weights for each case. 
For simplicity we work under the flat sky approximation throughout.
\subsection{Lensed fields and linear responses}
The lensed CMB temperature  $\tilde{T}_{\bm{\ell}}$ is related to the unlensed field $T_{\bm{\ell}}$ \textit{to linear order} in the lensing convergence $\kappa$ as follows \cite{2006PhR...429....1L}:
\beq
\label{eq:lensed_T}
\tilde{T}_{\bm{\ell}}
=
T_{\bm{\ell}}
-
2\int_{\bm{\ell}'} \frac{\bm{\ell}'\cdot (\bm{\ell}-\bm{\ell}')}{(\bm{\ell}-\bm{\ell}')^2} \kappa_{\bm{\ell}-\bm{\ell}'} T_{\bm{\ell}'}, 
\eeq
where $\int_{\bm{\ell}_1\cdots\bm{\ell}_n}\equiv \int d^2 \bm{\ell}_1\cdots d^2\bm{\ell}_n/(2\pi)^{2n}$. Throughout we use tildes to distinguish lensed fields from their unlensed counterparts. The analogous expression for the lensed polarization fields is:
\beq
\label{eq:lensed_EB}
\bal
\tilde{E}_{\bm{\ell}}\pm i \tilde{B}_{\bm{\ell}}
=E_{\bm{\ell}}
-2\int_{\bm{\ell}'} \frac{\bm{\ell}'\cdot (\bm{\ell}-\bm{\ell}')}{(\bm{\ell}-\bm{\ell}')^2}e^{\pm 2i (\phi_{\bm{\ell}'} - \phi_{\bm{\ell}})}
\kappa_{\bm{\ell}-\bm{\ell}'} E_{\bm{\ell}'},
\eal
\eeq 
where the angle $\phi_{\bm{\ell}}$ is defined through $\bm{\ell} = \ell \cos(\phi_{\bm{\ell}})\hat{\bm{x}}_1 + \ell \sin(\phi_{\bm{\ell}}) \hat{\bm{x}}_2$, $\{\hat{\bm{x}}_1,\hat{\bm{x}}_2\}$ are orthonormal basis vectors, and we've assumed that there are no primordial $B-$modes.

The lensed CMB temperature and polarization fields are statistically isotropic, and as a result $\langle \tilde{X}_{\bm{\ell}} \tilde{Y}_{\bm{L}-\bm{\ell}}\rangle =0$ for $L > 0$, where $\langle\cdots\rangle$ denotes an average over realizations of the primary CMB \textit{and} the lensing field, and $X, Y \in \{T, E, B\}$. However, fixing the Fourier mode $\kappa_{\bm{L}}$ breaks statistical isotropy, resulting in off diagonal covariances of the form:
\beq
\label{eq:response_definition}
\langle \tilde{X}_{\bm{\ell}} \tilde{Y}_{\bm{L}-\bm{\ell}} \rangle'
=
f^{XY}_{\bm{\ell},\bm{L}-\bm{\ell}}\,\,\kappa_{\bm{L}} + \mathcal{O}(\kappa^2)
\quad
\text{for}
\quad
L>0,
\eeq
where $f^{XY}_{\bm{\ell},\bm{L}-\bm{\ell}}$ is the linear response of lensed $\tilde{X},\tilde{Y}$ fields to the lensing convergence, and $\langle\cdots\rangle'$ denotes an average over realizations of the primary CMB and all Fourier modes of the lensing field \textit{except}\footnote{We could have also defined $\langle \cdots \rangle'$ as an average over the primary CMB with the entire lensing field fixed (not just the $\bm{L}$'th Fourier mode) and Eq.~\eqref{eq:response_definition} would still be valid. In \ref{sec:bias_hardening} our definition of only fixing the $\bm{L}$'th Fourier mode will be relevant for constructing the foreground QE, and we choose to adopt a common notation throughout for convenience.} for $\bm{L}$. 
Note that the responses trivially satisfy $f^{XY}_{\bm{\ell},\bm{\ell}'} = f^{YX}_{\bm{\ell}',\bm{\ell}}$.

From Eqs.~\eqref{eq:lensed_T} and \eqref{eq:lensed_EB} one obtains\footnote{Note that $\phi_{-\bm{\ell}} \cong \phi_{\bm{\ell}}+n\pi$ for any odd integer $n$. Since all angles in Eq.~\eqref{eq:responses} are multiplied by $2$, we are always free to change the sign of $\bm{\ell}$ in $\phi_{\bm{\ell}}$, which we have done to make the responses look as symmetric as possible.}:
\beq
\bal
\label{eq:responses}
%
f^{TT}_{\bm{\ell},\bm{L}-\bm{\ell}}
&=
\frac{2\bm{L}}{L^2}\cdot\left[\bm{\ell}C^{TT}_\ell + (\bm{L}-\bm{\ell})C^{TT}_{|\bm{L}-\bm{\ell}|}\right]\\
%
f^{TE}_{\bm{\ell},\bm{L}-\bm{\ell}}
&=
\frac{2\bm{L}}{L^2}\cdot\bigg[\bm{\ell}C^{TE}_\ell\cos(2\theta_{\bm{\ell},\bm{L}-\bm{\ell}})
+(\bm{L}-\bm{\ell})C^{TE}_{|\bm{L}-\bm{\ell}|}\bigg]\\
%
f^{TB}_{\bm{\ell},\bm{L}-\bm{\ell}}
&=
\frac{2\bm{L}}{L^2}\cdot\bm{\ell}C^{TE}_\ell\sin(2\theta_{\bm{\ell},\bm{L}-\bm{\ell}})\\
%
f^{EE}_{\bm{\ell},\bm{L}-\bm{\ell}}
&=
\frac{2\bm{L}}{L^2}\cdot\left[\bm{\ell}C^{EE}_\ell + (\bm{L}-\bm{\ell})
C^{EE}_{|\bm{L}-\bm{\ell}|}\right]
\cos(2\theta_{\bm{\ell},\bm{L}-\bm{\ell}})\\
%
f^{EB}_{\bm{\ell},\bm{L}-\bm{\ell}}
&=
\frac{2\bm{L}}{L^2}\cdot\bm{\ell}C^{EE}_\ell\sin(2\theta_{\bm{\ell},\bm{L}-\bm{\ell}})\\
%
f^{BB}_{\bm{\ell},\bm{L}-\bm{\ell}}
&=
0,
\eal
\eeq
where $C^{TT}_\ell, C^{TE}_\ell,$ and $C^{EE}_\ell$ are the unlensed power spectra, and $\theta_{\bm{\ell},\bm{\ell}'} \equiv \phi_{\bm{\ell}} - \phi_{\bm{\ell}'}$ is the angle between $\bm{\ell}$ and $\bm{\ell}'$.

\subsection{Minimum variance estimators}
\label{sec:MV}

Eq.~\eqref{eq:response_definition} is the starting point for deriving a quadratic estimator (QE) for the lensing convergence. By dividing both sides of Eq.~\eqref{eq:response_definition} by the linear response, one can build a simple unbiased QE $\hat{\kappa}^{XY}_{\bm{\ell},\bm{L}}$ for each $\bm{\ell}$: $\hat{\kappa}^{XY}_{\bm{\ell},\bm{L}} = \tilde{X}_{\bm{\ell}}\tilde{Y}_{\bm{L}-\bm{\ell}} / f^{XY}_{\bm{\ell},\bm{L}-\bm{\ell}}$, such that $\langle \hat{\kappa}^{XY}_{\bm{\ell},\bm{L}} \rangle' = \kappa_{\bm{L}}$ to linear order. The most general QE is obtained by taking an arbitrary weighted average of the $\hat{\kappa}^{XY}_{\bm{\ell},\bm{L}}$'s, provided that the weights sum to unity (to keep the estimator unbiased). 
Below we consider separately the specialized case where only a single pair of maps are used for the lensing reconstruction, and the more general scenario of summing over all available pairs. While the former is slightly suboptimal in terms of noise, it is still practically useful to have a collection of individual QEs for performing cross-checks and diagnosing systematics.

\subsubsection{Single pair of maps}
\label{sec:single_pair}
Given a single pair of maps $\tilde{X},\tilde{Y}$ one can construct a generic unbiased QE for the lensing convergence: 
\beq
\bal
\label{eq:kappa_XY}
&\hat{\kappa}^{XY}_{\bm{L}} = N^{XY}_{\bm{L}} \int_{\bm{\ell}} F^{XY}_{\bm{\ell},\bm{L}-\bm{\ell}} \tilde{X}_{\bm{\ell}} \tilde{Y}_{\bm{L}-\bm{\ell}}\\
&\text{where}\quad (N^{XY}_{\bm{L}})^{-1} = \int_{\bm{\ell}} F^{XY}_{\bm{\ell},\bm{L}-\bm{\ell}} f^{XY}_{\bm{\ell},\bm{L}-\bm{\ell}},
\eal
\eeq
where $F^{XY}_{\bm{\ell},\bm{L}-\bm{\ell}}$ are some weights, which are in principle arbitrary. The Gaussian component of the variance of $\hat{\kappa}^{XY}$, defined such that $\langle \hat{\kappa}^{XY}_{\bm{L}} \hat{\kappa}^{XY}_{\bm{L}'}\rangle_c = (2\pi)^2 \delta^D_{\bm{L} + \bm{L}'}\mathcal{N}^{XY}_{\bm{L}}$, is given by
\beq
\bal
\label{eq:noise_XY}
\mathcal{N}^{XY}_{\bm{L}} = (N^{XY}_{\bm{L}})^2 \int_{\bm{\ell}} F^{XY}_{\bm{\ell},\bm{L}-\bm{\ell}}&\bigg[F^{XY}_{\bm{\ell},\bm{L}-\bm{\ell}} \tilde{C}^{XX}_\ell \tilde{C}^{YY}_{|\bm{L}-\bm{\ell}|} \\&+ F^{XY}_{\bm{L}-\bm{\ell},\bm{\ell}} \tilde{C}^{XY}_\ell \tilde{C}^{XY}_{|\bm{L}-\bm{\ell}|}\bigg],
\eal
\eeq
where $\tilde{C}^{XY}_{\ell}$ is the \textit{total observed} power spectrum, i.e. the power spectra of the \textit{lensed} fields, including instrument noise and foregrounds. In the equation above we have made the simplifying assumption that the weights are real, so that $F^{XY}_{\bm{\ell},\bm{L}-\bm{\ell}} = F^{XY}_{-\bm{\ell},\bm{\ell}-\bm{L}}$. From here on out we will always assume that the weights and linear responses are real.

A common and optimistic choice for the weights is to minimize $\mathcal{N}^{XY}_{\bm{L}}$. This can be accomplished using e.g. functional derivatives, giving \cite{2002ApJ...574..566H}:
\beq
F^{XY}_{\bm{\ell},\bm{L}-\bm{\ell}}
=
\frac{\tilde{C}^{YY}_\ell \tilde{C}^{XX}_{|\bm{L}-\bm{\ell}|} f^{XY}_{\bm{\ell},\bm{L}-\bm{\ell}} - \tilde{C}^{XY}_\ell \tilde{C}^{XY}_{|\bm{L}-\bm{\ell}|} f^{XY}_{\bm{L}-\bm{\ell},\bm{\ell}}}{
\tilde{C}^{XX}_\ell \tilde{C}^{YY}_{|\bm{L}-\bm{\ell}|}\tilde{C}^{YY}_\ell \tilde{C}^{XX}_{|\bm{L}-\bm{\ell}|}
-\left(\tilde{C}^{XY}_\ell \tilde{C}^{XY}_{|\bm{L}-\bm{\ell}|}\right)^2
}.
\eeq
Note that in the limit where $\tilde{X}$ and $\tilde{Y}$ are uncorrelated (or when $\tilde{X} = \tilde{Y}$) the minimum variance weights reduce to $F^{XY}_{\bm{\ell},\bm{L}-\bm{\ell}} \propto f^{XY}_{\bm{\ell},\bm{L}-\bm{\ell}}/\tilde{C}^{XX}_\ell \tilde{C}^{YY}_{|\bm{L}-\bm{\ell}|}$ up to some generic prefactor (any function of $\bm{L}$) which the QE (Eq.~\eqref{eq:kappa_XY}) is insensitive to. 
Neglecting the cross-correlation between e.g. $T$ and $E$ is often a very good approximation, and doing so makes Eq.~\eqref{eq:kappa_XY} amenable to FFTs. Of course, this approximation to the weights still leaves the estimator unbiased, although slightly suboptimal.
In what follows we always make the approximation of neglecting the cross-correlation in the minimum variance weights for a single pair of maps.

\subsubsection{Summing over all pairs}
Given a collection of lensed maps $\{\tilde{M}^i\}$ one can build the general QE:
\beq
\bal
\label{eq:general_QE}
&\hat{\kappa}_{\bm{L}}
=
N^\kappa_{\bm{L}}
\sum_{ij}
\int_{\bm{\ell}}
F^{ij}_{\bm{\ell},\bm{L}-\bm{\ell}}
\tilde{M}^i_{\bm{\ell}}
\tilde{M}^j_{\bm{L}-\bm{\ell}}\\
&\text{where}\quad
(N^\kappa_{\bm{L}})^{-1} = 
\sum_{ij}
\int_{\bm{\ell}} 
F^{ij}_{\bm{\ell},\bm{L}-\bm{\ell}}
f^{ij}_{\bm{\ell},\bm{L}-\bm{\ell}},
\eal
\eeq
where the linear response $f^{ij}_{\bm{\ell},\bm{L}-\bm{\ell}}$ to the lensing convergence is defined in an analogous way as Eq.~\eqref{eq:response_definition}. Note that in Eq.~\eqref{eq:general_QE} we are summing over all pairs (e.g. both $ij$ and $ji$) and that the weights satisfy $F^{ij}_{\bm{\ell},\bm{\ell}'} = F^{ji}_{\bm{\ell}',\bm{\ell}}$. The index $i$ is quite general. For example, $i$ could simply index $T,E$ and $B$; or $\tilde{M}^i$ could be a temperature map with $i$ indexing the frequency; or even more generally $i$ could index a pair of indices $(X,\nu)$ where $X\in\{T,E,B\}$ and $\nu$ is a frequency. 

The variance of Eq.~\eqref{eq:general_QE} is:
\beq
\label{eq:global_noise}
\mathcal{N}^\kappa_{\bm{L}}
=
2 (N^\kappa_{\bm{L}} )^2
\sum_{ijmn}
\int_{\bm{\ell}}
F^{ij}_{\bm{\ell},\bm{L}-\bm{\ell}}
F^{mn}_{\bm{\ell},\bm{L}-\bm{\ell}}
\tilde{C}^{im}_\ell
\tilde{C}^{jn}_{|\bm{L}-\bm{\ell}|},
\eeq
where the covariance matrix $\tilde{C}^{ij}_\ell$ is defined as $\langle \tilde{M}^i_{\bm{\ell}} \tilde{M}^j_{\bm{L}-\bm{\ell}}\rangle = (2\pi)^2 \delta^D_{\bm{L}} \tilde{C}^{ij}_\ell$. 

The weights which minimize $\mathcal{N}^\kappa_{\bm{L}}$ are given by:
\beq
\label{eq:MV_weights}
F^{ij}_{\bm{\ell},\bm{L}-\bm{\ell}}
=
\frac{1}{2}
\sum_{mn}
\left(\tilde{\bm{C}}^{-1}_\ell\right)^{im} \left(\tilde{\bm{C}}^{-1}_{|\bm{L}-\bm{\ell}|}\right)^{jn} 
f^{mn}_{\bm{\ell},\bm{L}-\bm{\ell}},
\eeq
where we have chosen the normalization of the weights so that $\mathcal{N}^\kappa_{\bm{L}} = N^\kappa_{\bm{L}}$. 
Note that unlike for the single pair case, here the true minimum variance QE is always FFT-able. We provide explicit expressions for these weights in Appendix \ref{sec:GMV_explicit}.

Note that in the case where $\tilde{M}^i$ is a temperature map with $i$ indexing a frequency, Eq.~\eqref{eq:MV_weights} reduces to the familiar form: $w^i_\ell w^j_{|\bm{L}-\bm{\ell}|}f^{TT}_{\bm{\ell},\bm{L}-\bm{\ell}}/2\tilde{C}^\text{ILC}_\ell \tilde{C}^\text{ILC}_{|\bm{L}-\bm{\ell}|}$, where $\tilde{C}^\text{ILC}_\ell$ is the power spectrum of the minimum variance internal linear combination (MVILC) of temperature maps, and $w^i_\ell$ are the MVILC weights. In the case where $i$ indexes $T$, $E$, and $B$ (at a single frequency) Eq.~\eqref{eq:MV_weights} recovers the global minimum variance estimator (GMV), originally derived in \cite{Hirata:2003ka}. As was pointed out in \cite{Maniyar:2021msb}, the GMV has a slightly lower noise than the minimum variance linear combination of the MV single-pair estimators $\hat{\kappa}^{XY}_{\bm{L}}$.

\section{Bias hardened estimators}
\label{sec:bias_hardening}

The most general QE takes the form of Eq.~\eqref{eq:general_QE}, which is unbiased when the maps $\tilde{M}^i$ are composed of a lensed CMB signal and Gaussian noise, so that 
$
\langle \tilde{M}^i_{\bm{\ell}} \tilde{M}^j_{\bm{L}-\bm{\ell}}\rangle' = f^{ij}_{\bm{\ell},\bm{L}-\bm{\ell}}\,\kappa_{\bm{L}}. 
$
If instead the maps are contaminated by a foreground $s^i$, the covariance $\langle \tilde{M}^i_{\bm{\ell}} \tilde{M}^j_{\bm{L}-\bm{\ell}}\rangle'$ picks up additional off-diagonal contributions, which to lowest order in $\kappa$ and $s^i$ take the form\footnote{Note that in the average $\langle \cdots \rangle'$ we are fixing the $\bm{L}$'th Fourier mode of both the lensing convergence and the foreground.} \cite{2014JCAP...03..024O, Sailer:2020lal}:
\beq
\label{eq:biases}
\langle \tilde{M}^i_{\bm{\ell}} \tilde{M}^j_{\bm{L}-\bm{\ell}}\rangle' = f^{ij}_{\bm{\ell},\bm{L}-\bm{\ell}}\,\kappa_{\bm{L}} + \sum_k g^{ij,k}_{\bm{\ell},\bm{L}-\bm{\ell}}\,s^k_{\bm{L}},
\eeq
where as before $f^{ij}$ is the linear response to lensing in the absence of foregrounds, while $g^{ij,k}$ is the linear response to the foreground in the absence of lensing\footnote{Note that one could equivalently define $g^{ij,k}$ as the linear response of $\langle\tilde{M}^i \tilde{M}^j\rangle'$ to $s^k$ in the absence of lensing.}: $\langle s^i_{\bm{\ell}} s^j_{\bm{L}-\bm{\ell}}\rangle' = \sum_k g^{ij,k}_{\bm{\ell},\bm{L}-\bm{\ell}}\,s^k_{\bm{L}}$. The additional term in Eq.~\eqref{eq:biases} induces an additive bias to Eq.~\eqref{eq:general_QE} which is proportional to the foreground:
\beq
\label{eq:field_level_bias}
\langle \hat{\kappa}_{\bm{L}} \rangle'
=
\kappa_{\bm{L}}
+
\sum_{k}
\underbrace{
\left(
N^\kappa_{\bm{L}} \sum_{ij}\int_{\bm{\ell}} F^{ij}_{\bm{\ell},\bm{L}-\bm{\ell}} g^{ij,k}_{\bm{\ell},\bm{L}-\bm{\ell}}
\right)
}_{\equiv\, \mathcal{R}^{\kappa,k}_{\bm{L}}}
s^k_{\bm{L}},
\eeq
where we have defined the response $\mathcal{R}^{\kappa,i}_{\bm{L}}$ of the lensing estimator to the foreground in the $i$'th map.

The intuition behind bias hardening goes as follows: just as we can construct a convergence QE $\hat{\kappa}$ we can also construct a foreground QE $\hat{s}^i$ which can in turn be used to subtract off the bias to Eq.~\eqref{eq:field_level_bias}. The only ingredients that we need are the foreground linear response and a set of weights $G^{ij,k}$:
\beq
\bal
&\hat{s}^i_{\bm{L}} = N^i_{\bm{L}} \sum_{jk} \int_{\bm{\ell}} G^{jk,i}_{\bm{\ell},\bm{L}-\bm{\ell}} \tilde{M}^j_{\bm{\ell}} \tilde{M}^k_{\bm{L}-\bm{\ell}}
\\
&\text{where}\quad
(N^i_{\bm{L}})^{-1} = \sum_{jk}\int_{\bm{\ell}} G^{jk,i}_{\bm{\ell},\bm{L}-\bm{\ell}} g^{jk,i}_{\bm{\ell},\bm{L}-\bm{\ell}}.
\eal
\eeq
When evaluated on the contaminated maps, $\hat{s}^i$ picks up field level biases proportional to the foreground in the other maps $(m\neq i)$, and to $\kappa$:
\beq
\bal
\label{eq:biases_pt2}
\langle \hat{s}^i_{\bm{L}} \rangle'
&=
\sum_m
\left(
N^i_{\bm{L}}
\sum_{jk}\int_{\bm{\ell}}
G^{jk,i}_{\bm{\ell},\bm{L}-\bm{\ell}}
g^{jk,m}_{\bm{\ell},\bm{L}-\bm{\ell}}
\right)
s^m_{\bm{L}}\\
&\,\,\,\,\,\,\,\,\,\,
+
\left(
N^i_{\bm{L}}
\sum_{jk}
\int_{\bm{\ell}}
G^{jk,i}_{\bm{\ell},\bm{L}-\bm{\ell}}
f^{jk}_{\bm{\ell},\bm{L}-\bm{\ell}}
\right)
\kappa_{\bm{L}}\\
&\equiv
\sum_m \mathcal{R}^{i,m}_{\bm{L}} s^m_{\bm{L}}
+
\mathcal{R}^{i,\kappa}_{\bm{L}} \kappa_{\bm{L}},
\eal
\eeq
where $\mathcal{R}^{i,i}_{\bm{L}} = 1$. By recasting Eqs.~\eqref{eq:field_level_bias} and \eqref{eq:biases_pt2} in a matrix form one can construct a suitable linear combination of $\hat{\kappa}$ and $\hat{s}^i$ to null the linear order field level biases:
\beq
\label{eq:bias_hardened}
\begin{pmatrix}
\hat{\kappa}^\text{BH}_{\bm{L}} \\
\hat{s}^{\text{BH},1}_{\bm{L}} \\
\vdots \\
\hat{s}^{\text{BH},n}_{\bm{L}} \\
\end{pmatrix}
=
\begin{pmatrix}
1 & \mathcal{R}^{\kappa,1}_{\bm{L}} & \cdots & \mathcal{R}^{\kappa,n}_{\bm{L}}\\
\mathcal{R}^{1,\kappa}_{\bm{L}} & 1 & \cdots & \mathcal{R}^{1,n}_{\bm{L}} \\
\vdots & \vdots & \ddots & \vdots \\
\mathcal{R}^{n,\kappa}_{\bm{L}} & \mathcal{R}^{n,1}_{\bm{L}} & \cdots & 1
\end{pmatrix}^{-1}
\begin{pmatrix}
\hat{\kappa}_{\bm{L}} \\
\hat{s}^1_{\bm{L}} \\
\vdots \\
\hat{s}^n_{\bm{L}} \\
\end{pmatrix}.
\eeq
By taking an average of Eq.~\eqref{eq:bias_hardened} and plugging in Eqs.~\eqref{eq:field_level_bias} and \eqref{eq:biases_pt2}, we see that the bias hardened estimators satisfy: $\langle \hat{\kappa}^\text{BH}_{\bm{L}}\rangle' = \kappa_{\bm{L}}$ and $\langle \hat{s}^{\text{BH},i}_{\bm{L}}\rangle' = s^i_{\bm{L}}$.

In practice we choose the foreground weights $G$ and lensing weights $F$ to minimize the variance of $\hat{s}^i$ and $\hat{\kappa}$ respectively. In Appendix \ref{sec:optimal_bias_hardening} we show that this choice of weights is equivalent to the minimum variance lensing estimator constrained to have zero response to foregrounds (i.e. the optimal bias hardened estimator).  

For the case of using a single pair of maps for reconstruction we therefore take $G^{ij,k}_{\bm{\ell},\bm{L}-\bm{\ell}} = g^{ij,k}_{\bm{\ell},\bm{L}-\bm{\ell}}/\tilde{C}^{ii}_\ell \tilde{C}^{jj}_{|\bm{L}-\bm{\ell}|}$ and $F^{ij}_{\bm{\ell},\bm{L}-\bm{\ell}} = f^{ij}_{\bm{\ell},\bm{L}-\bm{\ell}}/\tilde{C}^{ii}_\ell \tilde{C}^{jj}_{|\bm{L}-\bm{\ell}|}$, where we have neglected the cross-correlation $\tilde{C}^{ij}$ to make the estimators FFT-able. As an illustrative example we explicitly bias harden the $TE$ estimator in Appendix \ref{sec:explicit_te_psh} using the foreground linear responses derived in section \ref{sec:foreground_linear_response}.

\subsection{Foreground model}
\label{sec:foreground_model}

The only input required to construct the foreground QE is the linear response. Below we outline our model for the foregrounds. We use this model to compute the linear response in the following subsection.

We assume that the foreground $s$ is a sum of $N_\text{sources}$ sources with identical profiles, and further assume that polarization fractions and polarization angles do not vary significantly over a single source, so that the intensity and polarization of the foreground take the form:
\beq
\bal
\label{eq:IQU_sources}
s^I(\bm{x}) &= \sum_i s^I_i u(\bm{x}-\bm{x}_i)\\
s^Q(\bm{x}) \pm i s^U(\bm{x}) &= \sum_i s^I_i p_i e^{\pm 2i \psi_i} u(\bm{x}-\bm{x}_i)
\eal
\eeq
where $u$ is the profile, and $s^I_i, p_i, \psi_i, \bm{x}_i$ is the flux, polarization fraction, polarization angle, and position of the $i$'th source respectively. 
For simplicity we will assume that the sources emit no circular polarization\footnote{While the primary CMB is expected to have a negligible level of circular polarization in absence of new physics \cite{Montero-Camacho:2018vgs}, physical processes such as synchrotron emission (which is responsible for radio point sources) can have a circularly-polarized component, which may be potentially detectable by future experiments at low frequencies ($\nu \lesssim 10$ GHz) \cite{King:2016exc}. It appears that circular polarization should be negligible at CMB frequencies, and it will not be explored further in this work.}. 
From the definition of the $E$ and $B$ fields \cite{Zaldarriaga:1996xe,2006PhR...429....1L} 
\beq
E_{\bm{\ell}} \pm i B_{\bm{\ell}}
= -\big[Q_{\bm{\ell}} \pm i U_{\bm{\ell}}\big] e^{\mp 2i\phi_{\bm{\ell}}}
\eeq
the $T,E,$ and $B$ fields for the sources take the form:
\beq
\label{eq:foregrounds_fourier}
\bal
s^T_{\bm{\ell}} &= \sum_i s_i e^{-i\bm{\ell}\cdot\bm{x}_i}u_{\bm{\ell}}\\
s^E_{\bm{\ell}} \pm i s^{B}_{\bm{\ell}} &= - \sum_i s_i p_i e^{\pm 2i(\psi_i-\phi_{\bm{\ell}})} e^{-i \bm{\ell}\cdot\bm{x}_i} u_{\bm{\ell}},
\eal
\eeq
where $s_i$ is the flux in the appropriate units.

To compute the responses we must make some assumptions about the joint PDF of the sources $P(\{s_i, p_i, \psi_i, \bm{x}_i \})$. As we will see in the next subsection, this is easily accomplished with the following simple set of assumptions:
\begin{itemize}
    \item Each source is statistically independent. 
    \item Positions and polarization angles are statistically independent, both of each other, and independent of polarization fractions and fluxes.
    \item Positions and polarization angles are uniformly distributed.
    This key assumption neglects the clustering of sources, and is appropriate for the polarized radio and infrared sources we consider here (on the small scales relevant for lensing reconstruction).
\end{itemize}
These assumptions force the joint PDF to factorize as\footnote{In practice, we do not need to assume a specific distribution for the number of sources $N_\text{sources}$ to compute the response. However, as soon as the sources are independent, the number of sources in any subpatch of the map will be Poisson distributed.}:
\beq
\bal
\label{eq:PDF}
&P\left( \{s_i, p_i, \psi_i, \bm{x}_i \}\right)
=
P\left( N_\text{sources} \right)\\
&\times\prod_{i=1}^{N_\text{sources}}
\mathcal{U}\left( \bm{x}_i \right)
\mathcal{U}\left( \psi_i \right)
\text{LF}\left( s_i \right)
P\left( p_i |s_i \right).
\eal
\eeq
Here, ``LF" is the source luminosity function, normalized to integrate to unity, and $\mathcal{U}$ denotes a uniform distribution (in coordinate or angle).

Note that by forcing each source to be independent and uniformly distributed on the sky, any $n$-point function of the foreground will only involve one independent sum over sources, since
\begin{equation*}
\langle \text{Exp}[-i(\bm{\ell}_{1}\cdot\bm{x}_{i_1} + \cdots + \bm{\ell}_n \cdot \bm{x}_{i_n})] \rangle 
\end{equation*}
vanishes unless $i_1 = \cdots = i_n$ and $\sum_i \bm{\ell}_i = 0$, assuming that no non-trivial subset of the $\bm{\ell}_i$'s sum to zero, which will always be the case for the responses below. That is, our assumptions create the usual shot noise $n$-point functions. 

\subsection{Foreground linear response}
\label{sec:foreground_linear_response}

Our goal is to compute the linear response $g^{XY,Z}_{\bm{\ell},\bm{L}-\bm{\ell}}$ of the covariance $\langle s^X_{\bm{\ell}} s^Y_{\bm{L}-\bm{\ell}}\rangle'$ to the field $s^Z_{\bm{L}}$ using the simple foreground model outlined in the previous subsection. 

We first note that the source cross-power spectra of $T$, $E$ and $B$ vanish, because the polarization angles are uniform and independent:
\beq
\label{eq:orthogonal}
\langle s^X_{\bm{L}} s^Y_{-\bm{L}} \rangle \propto \delta^K_{XY}.
\eeq 
A more rigorous argument for this orthogonality goes as follows: when $X=T$ and $Y=E$ or $B$, the LHS of Eq.~\eqref{eq:orthogonal} vanishes since $\langle e^{\pm 2 i \psi_i}\rangle=0$. When $X=E$ and $Y=B$, we average over $\langle \cos(2\psi_i - 2\phi_{\bm{L}}) \sin(2\psi_i - 2\phi_{-\bm{L}})\rangle \propto \sin(2\phi_{\bm{L}}-2\phi_{-\bm{L}})=0$. Eq.~\eqref{eq:orthogonal} follows from these two observations.

With Eq.~\eqref{eq:orthogonal} at our disposal, it is straightforward to compute the responses.  
We start with the definition of the response $g$: $\langle s^X_{\bm{\ell}} s^Y_{\bm{L}-\bm{\ell}}\rangle' = \sum_Z g^{XY,Z}_{\bm{\ell},\bm{L}-\bm{\ell}} \,\,s^Z_{\bm{L}}$, we multiply both sides by $s_{-\bm{L}}^{Z'}$, and take an average:
\beq
g^{XY,Z}_{\bm{\ell},\bm{L}-\bm{\ell}} 
=
\frac{\langle s^X_{\bm{\ell}} s^Y_{\bm{L}-\bm{\ell}} s^Z_{-\bm{L}}\rangle}{\langle s^Z_{\bm{L}}s^Z_{-\bm{L}}\rangle}.
\eeq

Using this expression and the foreground PDF (Eq.~\eqref{eq:PDF}), we can explicitly compute all unique $18$ responses, which are given in Table~\ref{tab:response}.
Similar responses to $s^Q$ and $s^U$ were recently derived in Appendix E of \cite{BICEPKeck:2022kci}.
\begin{table}[!h]
\resizebox{\columnwidth}{!}{
\begin{tabular}{c||c|c|c}
      & $Z = T$ & $E$ & $B$\\
     \hline
     \hline
     $XY = TT$ & $\mathcal{A}$ & 0 & 0\\
     \hline
     $TE$ & 0 & $\mathcal{B} \cos(2\theta_{\bm{L},\bm{L}-\bm{\ell}})$ & $\mathcal{B} \sin(2\theta_{\bm{L}-\bm{\ell},\bm{L}})$\\
     \hline
     $TB$ & 0 & $\mathcal{B} \sin(2\theta_{\bm{L},\bm{L}-\bm{\ell}})$ &  $\mathcal{B} \cos(2\theta_{\bm{L}-\bm{\ell},\bm{L}})$\\
     \hline
     $EE$ & $\mathcal{C} \cos(2\theta_{\bm{\ell},\bm{L}-\bm{\ell}})$ & 0 & 0\\
     \hline
     $EB$ & $\mathcal{C} \sin(2\theta_{\bm{\ell},\bm{L}-\bm{\ell}})$ & 0 & 0\\
     \hline
     $BB$ & $\mathcal{C} \cos(2\theta_{\bm{\ell},\bm{L}-\bm{\ell}})$ & 0 & 0
\end{tabular}}
\caption{The linear response $g^{XY,Z}_{\bm{\ell},\bm{L}-\bm{\ell}}$ for all unique combinations of $XY$ (rows) and $Z$ (columns). Note that all entries in the table should be multiplied by a geometric factor $u_{\bm{\ell}}u_{\bm{L}-\bm{\ell}}/u_{\bm{L}}$. 
For brevity we have defined
$\mathcal{A}\equiv \langle s_i^3\rangle/\langle s_i^2\rangle$,
$\mathcal{B}\equiv \langle p_i^2 s^3_i\rangle / \langle p_i^2 s_i^2\rangle $, and
$\mathcal{C}\equiv \langle p_i^2 s^3_i\rangle / 2\langle s_i^2\rangle$.
}
\label{tab:response}
\end{table}

As an illustrative example we will calculate the response of $TE$ to $B$ for the specialized case of point sources ($u_{\bm{\ell}} = 1$). 
We first compute the bispectrum $\langle s^T_{\bm{\ell}} s^{E}_{\bm{L}-\bm{\ell}} s^B_{-\bm{L}} \rangle$, given by:
\beq
\bal
\label{eq:EQ1}
&\sum_i \langle s_i^3p_i^2\rangle 
\int \frac{d\psi_i}{2\pi}
\cos(2(\psi_i-\phi_{\bm{L}-\bm{\ell}}))
\sin(2(\psi_i-\phi_{-\bm{L}})) 
\\
&=
\frac{\langle N_\text{sources} \rangle }{2}
\langle s_i^3p_i^2\rangle
 \underbrace{\sin(2(\phi_{\bm{L}-\bm{\ell}} - \phi_{-\bm{L}}))}_{=\,\sin(2\theta_{\bm{L}-\bm{\ell},\bm{L}})}.
\eal
\eeq
A similar calculation gives the power spectrum $\langle s^B_{\bm{L}} s^B_{-\bm{L}}\rangle$:
\beq
\bal
\label{eq:EQ2}
&\sum_i \langle s_i^2p_i^2\rangle 
\int \frac{d\psi_i}{2\pi}
\sin(2(\psi_i-\phi_{\bm{L}}))
\sin(2(\psi_i-\phi_{-\bm{L}})) 
\\
&=
\frac{\langle N_\text{sources} \rangle }{2} 
\langle s_i^2p_i^2\rangle
\underbrace{\cos(2(\phi_{\bm{L}} - \phi_{-\bm{L}}))}_{=\,1}.
\eal
\eeq
The response $g^{TE,B}_{\bm{\ell},\bm{L}-\bm{\ell}}$ is given by the ratio of Eq.~\eqref{eq:EQ1} to Eq.~\eqref{eq:EQ2}, corresponding to the value reported in Table \ref{tab:response}. Note that the factor $\langle N_\text{sources}\rangle$ cancels, which is why we do not explicitly need to assume the distribution of the number of sources.

\subsection{Picking the wrong profile}
\label{sec:wrong_profile}

Here we explore the expected bias to a bias hardened estimator (hardened against a profile $u_{\bm{\ell}}$) from a foreground with profile $w_{\bm{\ell}} \neq u_{\bm{\ell}}$. For brevity we will consider the case of temperature-only reconstruction in the large-lens limit ($L\to 0$). 

For the case of temperature-only reconstruction the optimal bias hardened lensing weights take the form:
\beq
\label{eq:bias_hardened_TT}
F^\text{BH}_{\bm{\ell},\bm{L}-\bm{\ell}} 
=
F^\text{MV}_{\bm{\ell},\bm{L}-\bm{\ell}} 
-
\frac{\int_{\bm{\ell}'} F^\text{MV}_{\bm{\ell}',\bm{L}-\bm{\ell}'}  u_{\ell'} u_{|\bm{L}-\bm{\ell}'|} }{
\int_{\bm{\ell}'} G^\text{MV}_{\bm{\ell}',\bm{L}-\bm{\ell}'} 
u_{\ell'} u_{|\bm{L}-\bm{\ell}'|}} 
G^\text{MV}_{\bm{\ell},\bm{L}-\bm{\ell}},
\eeq
as can be shown with Eq.~\eqref{eq:bias_hardened}.
In Eq.~\eqref{eq:bias_hardened_TT} we have assumed that the input profile $u_\ell$ is isotropic, and $F^\text{MV}\,(G^\text{MV})$ are the minimum variance temperature-only weights for the lensing convergence (foreground):
\beq
\bal
\label{eq:MV_temperature_weights}
F^\text{MV}_{\bm{\ell},\bm{L}-\bm{\ell}} &= f^{TT}_{\bm{\ell},\bm{L}-\bm{\ell}} / 2 \tilde{C}^{TT}_\ell C^{TT}_{|\bm{L}-\bm{\ell}|}\\
G^\text{MV}_{\bm{\ell},\bm{L}-\bm{\ell}} &= \mathcal{A} u_\ell u_{|\bm{L}-\bm{\ell}|} / u_L 2 \tilde{C}^{TT}_\ell C^{TT}_{|\bm{L}-\bm{\ell}|}.
\eal
\eeq

As an aside, note that the bias hardened weights (Eq.~\eqref{eq:bias_hardened_TT}) are insensitive to an overall normalization of $G^\text{MV}$. Hence bias hardening the $TT$ estimator requires no knowledge of the source amplitude $\mathcal{A}$, or the normalization of the profile. In Appendix \ref{sec:GPSH} we show that this is always the case when bias hardening any single-pair lensing estimator, but is no longer true when bias hardening the global minimum variance estimator.

Suppose that the true foreground has an isotropic profile $w_\ell\neq u_\ell$. In the $L\to 0$ limit the response of a generic lensing estimator $\hat{\kappa}$ (with temperature weights $F^\kappa$ and normalization $N^\kappa$) to the true foreground is:
\beq
\bal
\mathcal{R}^{\kappa,w}_{\bm{0}}&=
N^\kappa_{\bm{0}} \int_{\bm{\ell}} F^\kappa_{\bm{\ell},-\bm{\ell}}\, g^{TT,T}_{\bm{\ell},-\bm{\ell}}\\
&=
\mathcal{A}
\int \frac{d\ell}{(2\pi)^2} w^2_\ell
\left[
N^\kappa_{\bm{0}} \ell
\int d\theta_{\bm{L},\bm{\ell}}
F^\kappa_{\bm{\ell},-\bm{\ell}} 
\right],
\eal
\eeq
such that the field level bias to $\langle\hat{\kappa}_{\bm{0}}\rangle'$ is $\mathcal{R}^{\kappa,w}_{\bm{0}} s^T_{\bm{0}}$. Here we have assumed that the profile is normalized so that $\lim_{\ell\to 0}u_\ell =1$.

\begin{figure}[!h]
\centering
\includegraphics[width=0.9\linewidth]{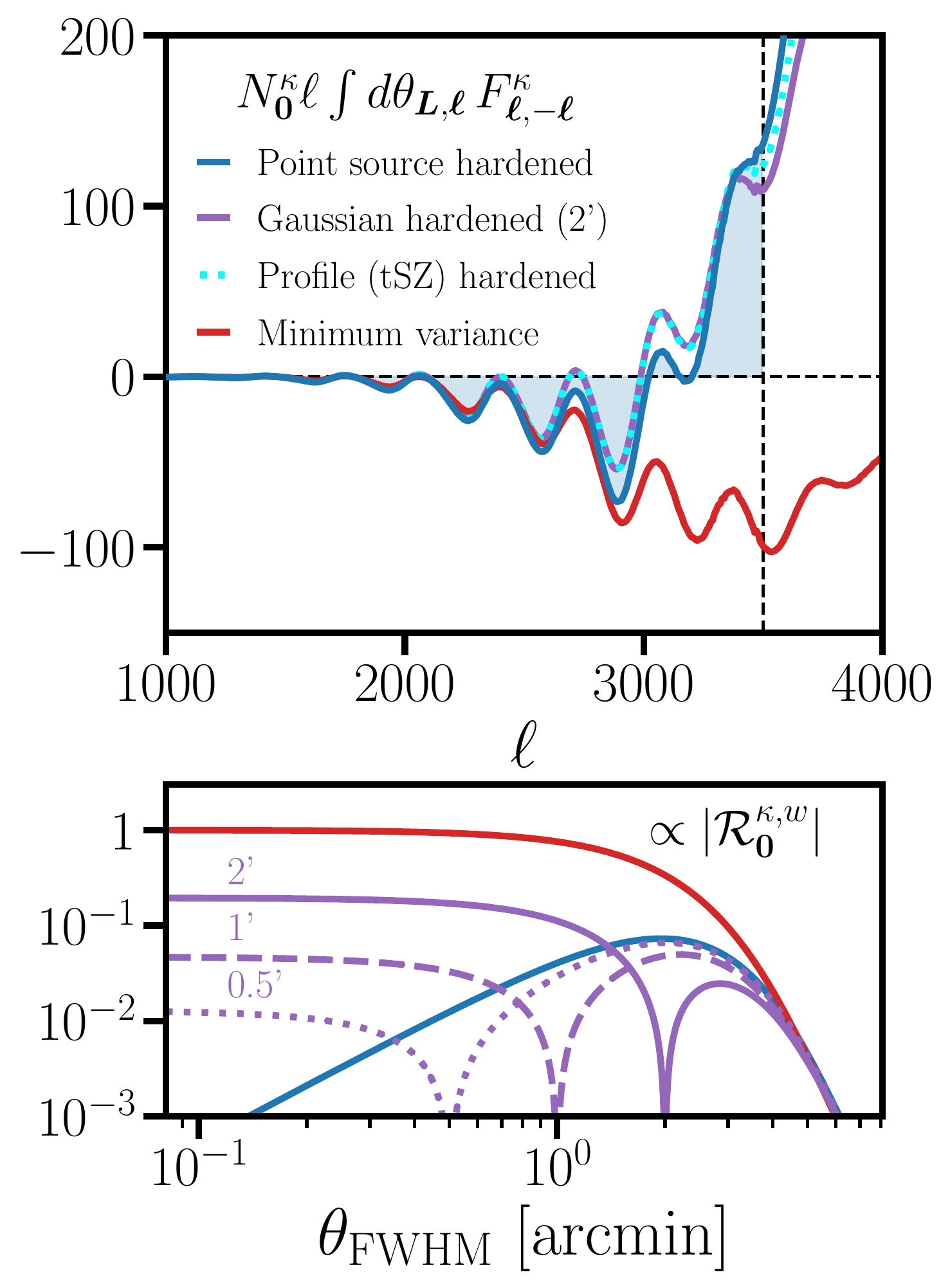}
\caption{ \textit{Top:} The angular averaged temperature-only lensing weights ($\times\,2\pi N^\kappa_{\bm{0}}\ell$) in the large-lens limit $(L\to0)$ when hardening against a point source (blue), hardening against a Gaussian profile with a $2'$ FHWM (purple) and when choosing the weights to minimize the variance (red). Note that since the point source hardened estimator has zero response to point-sources (when $w_\ell = 1$) the area under the blue curve vanishes. \textit{Bottom:} The response to a Gaussian foreground $w_\ell$ with a full width half max $\theta_{\text{FHHM}}$, normalized to the response of the MV estimator to point sources. In both plots we take $\ell_{\text{max},T}=3500$. See section \ref{sec:noise} for details regarding the instrumental assumptions.
}
\label{fig:wrong_profile}
\end{figure}

In the bottom panel of Fig.~\ref{fig:wrong_profile} we show the values of $|\mathcal{R}^{\kappa,w}_{\bm{0}}|$ when the true foreground profile $w_\ell$ is a Gaussian with a full width half maximum (FWHM) given by $\theta_\text{FWHM}$ (in arcmin), and consider three cases for $F^\kappa$: minimum variance weights (red), hardening against a Gaussian profile (purple), and hardening against a point source (blue). The angular averaged values of these weights are plotted in the top panel of Fig.~\ref{fig:wrong_profile}.

We find that point source hardening reduces the response to foregrounds smaller than $\sim 2$ arcmin by an order of magnitude relative to the MV weights. By instead hardening against a Gaussian with a non-zero width, one can further reduce (or precisely null) the response to foregrounds whose angular size is within $\sim 40\%$ of the size of the profile. In particular the tSZ-profile hardened estimator \cite{Sailer:2020lal} (cyan), which is well approximated by the Gaussian-hardened estimator with a 2 arcmin FWHM, reduces the response to point sources by roughly a factor of $5$ relative to the MV estimator. 

Finally, if the foreground is larger than a few arcmin, the response is always small regardless of the assumed profile or technique, since the primary CMB dominates dominates on these scales (i.e. the lensing weights have no support for $\ell < 1500$). 

\section{Large-lens suppressed estimators}
\label{sec:LLS}

In this section we study the ``large-lens'' or ``squeezed'' limit ($L\ll\ell$) analytically. A similar study for the temperature-only case has given rise to ``shear-only'' estimators \cite{2019PhRvL.122r1301S}, which are highly effective at reducing foreground biases in this limit. Here we extend this study to polarization, highlighting important differences compared to temperature. 

As in section \ref{sec:bias_hardening}, define $\mathcal{R}^{XY,Z}_{\bm{L}}$ as the response of a single-pair lensing estimator to a foreground in the $Z$'th map:
\beq
\label{eq:responseXY}
(N^{XY}_{\bm{L}})^{-1}\,\mathcal{R}^{XY,Z}_{\bm{L}} = 
\int \frac{\ell d\ell}{(2\pi)^2} \int d\theta_{\bm{L},\bm{\ell}} \, F^{XY}_{\bm{\ell},\bm{L}-\bm{\ell}} \,g^{XY,Z}_{\bm{\ell},\bm{L}-\bm{\ell}},
\eeq
so that the field level bias to $\langle \hat{\kappa}^{XY}_{\bm{L}}\rangle'$ due to $s^Z_{\bm{L}}$ is $\mathcal{R}^{XY,Z}_{\bm{L}} s^Z_{\bm{L}}$. The inverse of the normalization can similarly be expressed as: 
\beq
\label{eq:inverse_norm}
(N^{XY}_{\bm{L}})^{-1} = 
\int \frac{\ell d\ell}{(2\pi)^2} \int d\theta_{\bm{L},\bm{\ell}} \, F^{XY}_{\bm{\ell},\bm{L}-\bm{\ell}} f^{XY}_{\bm{\ell},\bm{L}-\bm{\ell}}.
\eeq
The large-lens suppressed (LLS) estimators are derived by choosing the weights to suppress the field level bias by $\mathcal{O}((L/\ell)^2)$ relative to the lensing signal, where $\ell\sim 3000$ is evaluated in the domain where the lensing estimator receives the majority of its signal (see the top panel of Fig.~\ref{fig:wrong_profile}). This can be accomplished by forcing Eq.~\eqref{eq:responseXY} to be suppressed by $\mathcal{O}((L/\ell)^2)$ while simultaneously retaining a $\mathcal{O}(1)$ scaling in the inverse normalization.

We first consider the temperature-only case. By Taylor expanding the angular-dependent piece of foreground linear response $g^{TT,T}_{\bm{\ell},\bm{L}-\bm{\ell}} = \mathcal{A} u_\ell u_{\bm{L}-\bm{\ell}} / u_L$ in powers of $L/\ell$ (see Appendix \ref{sec:taylor_expansions}) we find:
\beq
\bal
\label{eq:taylor_RTTT}
&g^{TT,T}_{\bm{\ell},\bm{L}-\bm{\ell}} 
= \mathcal{A}  \frac{u_\ell}{u_L}
\bigg\{
u_\ell - L u'_\ell \cos(\theta_{\bm{L},\bm{\ell}})
\\&+\frac{1}{4}\left(\frac{L}{\ell}\right)^2
\bigg[ D^+_\ell  + \cos(2\theta_{\bm{L},\bm{\ell}})D^-_\ell \bigg]u_\ell
+ \mathcal{O}\left((L/\ell)^3\right)
\bigg\}
,
\eal
\eeq
with corrections going as $\mathcal{O}\left((L/\ell)^3\right)$, and we have defined the operators $D^\pm_\ell \equiv \ell^2 d^2/d\ell^2 \pm d/d\ln\ell$. Likewise, the linear response of the CMB to lensing also factorizes into a nice multipole form \cite{2018JCAP...01..034P,2019PhRvL.122r1301S}:
\beq
\label{eq:taylor_fTT}
f^{TT}_{\bm{\ell},\bm{L}-\bm{\ell}}
= 
C^{TT}_\ell
\left[
\frac{d\ln \ell^2 C^{TT}_\ell}{d\ln\ell}
+
\cos(2\theta_{\bm{L},\bm{\ell}})
\frac{d\ln C^{TT}_\ell}{d\ln\ell}
\right]
\eeq
with corrections going as $\mathcal{O}(L/\ell)$.

In a similar manner one can expand the lensing weights into multipoles:
\beq
\bal
F^{TT}_{\bm{\ell},\bm{L}-\bm{\ell}} &= \sum_{m=0}^\infty \left(A^{(m)}_{L,\ell} \cos(m \theta_{\bm{L},\bm{\ell}}) + B^{(m)}_{L,\ell} \sin(m \theta_{\bm{L},\bm{\ell}}) \right),
\eal
\eeq
and in turn express the coefficients $A^{(m)}$ and $B^{(m)}$ as a Taylor expansion in $L/\ell$. From Eq.~\eqref{eq:taylor_RTTT}, we see that setting $A^{(0)} =\mathcal{O}((L/\ell)^2),\,A^{(1)} = \mathcal{O}(L/\ell)$ forces the response to foregrounds \eqref{eq:responseXY} to be suppressed by $\mathcal{O}((L/\ell)^2)$. 
From Eq.~\eqref{eq:taylor_fTT}, we see that retaining a $\mathcal{O}(1)$ scaling in Eq.~\eqref{eq:inverse_norm} requires choosing $A^{(2)} = \mathcal{O}(1)$. The remaining coefficients are chosen to minimize the noise of the LLS estimator in the large-lens limit. The large-lens noise (Eq.~\eqref{eq:noise_XY} as $L\to0$) of the temperature-only LLS estimator takes the form:
\beq
\bal
\label{eq:noise_large_lens}
\mathcal{N}^{TT}_0 = \frac{\int\frac{\ell d\ell}{2\pi} (C^{TT}_\ell)^2 \left[\sum_{m=2}^\infty \big(A^{(m)}_{0,\ell}\big)^2 + \sum_{m=1}^\infty\big(B^{(m)}_{0,\ell}\big)^2 \right]}{
\left[\int \frac{\ell d\ell}{4\pi} A^{(2)}_{0,\ell} \,\frac{d C^{TT}_\ell}{ d\ln\ell}\right]^2
}.
\eal
\eeq
Note that the $A^{(2)}$-term is the only term contributing to the normalization (since we set $A^{(0)}_{0,\ell}=0$). Thus a non-zero value for the remaining coefficients ($A^{(m)}$ for $m>2$ and all $B^{(m)}$) only increases the noise of the LLS estimator. To minimize the large-lens noise, we therefore choose the remaining coefficients to be $\mathcal{O}(L/\ell)$, and solve for the value of $A^{(2)}_{0,\ell}$ which minimizes Eq.~\eqref{eq:noise_large_lens}. Doing so forces the LLS estimator to take the form:
\beq
\label{eq:LLS_temp}
F^{\text{LLS},TT}_{\bm{\ell},\bm{L}-\bm{\ell}} = \frac{1}{2(\tilde{C}^{TT}_\ell)^2} \frac{d C^{TT}_\ell}{ d\ln\ell} \cos(2\theta_{\bm{L},\bm{\ell}}) + \mathcal{O}(L/\ell),
\eeq
recovering the shear-only estimator \cite{2019PhRvL.122r1301S}. As was shown in \cite{Qu:2022qie}, while the shear-only estimator suppresses the bias by $\mathcal{O}((L/\ell)^2)$, the bias \textit{is not precisely nulled} for a source with a generic profile, even if the profile is azimuthally symmetric. We further note that Eq.~\eqref{eq:taylor_RTTT} also contains derivatives of the (Fourier-transformed) profile, which additionally suppress biases for small sources. In the special case of point sources, for which these derivatives vanish, the shear-only estimator is non-perturbatively unbiased.

We explicitly generalize this procedure for all pairs of temperature and polarization maps in Appendix \ref{sec:LLS_details}. Our results are summarized by:
\beq
\bal
\label{eq:LLS_wpol}
F^{\text{LLS},TE}_{\bm{\ell},\bm{L}-\bm{\ell}} &= \frac{C^{TE}_\ell}{\tilde{C}^{TT}_\ell \tilde{C}^{EE}_\ell} \frac{d \ln\ell^2 |C^{TE}_\ell|}{d\ln\ell} + \mathcal{O}(L/\ell)\\
F^{\text{LLS},TB}_{\bm{\ell},\bm{L}-\bm{\ell}} &=\mathcal{O}(L/\ell)\\
F^{\text{LLS},EE}_{\bm{\ell},\bm{L}-\bm{\ell}} &= \frac{1}{2(\tilde{C}^{EE}_\ell)^2} \frac{d C^{EE}_\ell}{ d\ln\ell} \cos(2\theta_{\bm{L},\bm{\ell}}) + \mathcal{O}(L/\ell)\\
F^{\text{LLS},EB}_{\bm{\ell},\bm{L}-\bm{\ell}} &= F^{\text{MV},EB}_{\bm{\ell},\bm{L}-\bm{\ell}} =  \frac{f^{EB}_{\bm{\ell},\bm{L}-\bm{\ell}}}{\tilde{C}^{EE}_\ell \tilde{C}^{BB}_{|\bm{L}-\bm{\ell}|}}.
\eal
\eeq
One can continue this procedure to obtain LLS estimators to any order in $L/\ell$. Including higher order corrections will in principle make the estimators less noisy, however, by construction the approximations in Eqs.~\eqref{eq:LLS_temp} and \eqref{eq:LLS_wpol} are only slightly suboptimal in the regime where the large-lens expansion is valid. 

As indicated by Eq.~\eqref{eq:LLS_wpol}, we find that there is no leading order LLS $TB$ estimator. As shown in Appendix \ref{sec:taylor_expansions} and \ref{sec:LLS_details}, this is because $f^{TB} \propto g^{TB,E} \propto \sin(2\theta)$ to lowest order in the large-lens limit, i.e. foregrounds and lensing are indistinguishable to lower order.  On the contrary we find that the MV $EB$ estimator already satisfies the LLS requirements. 

\begin{table}[!h]
\begin{tabular}{c||c|c}
      & $\mathcal{O}(\text{MV biases})$ & $\mathcal{O}(\text{LLS suppression})$\\
     \hline
     \hline
     $TT$ & $1$ & $L^2 u''_\ell/u_\ell\,,\,L^2 u'_\ell/\ell u_\ell$ \\
     \hline
     $TE$ & $p_\text{source}/ p_\text{CMB}$ & $L^2 u''_\ell/u_\ell\,,\,L^2 u'_\ell/\ell u_\ell$ \\
     \hline
     $TB$ & $p_\text{source}/ p_\text{CMB}$ & $-$ \\
     \hline
     $EE$ & $(p_\text{source}/ p_\text{CMB})^2$ & $(L/\ell)^2$ \\
     \hline
     $EB$ & $(L/\ell)^2 \times (p_\text{source}/ p_\text{CMB})^2$ & $-$ 
\end{tabular}
\caption{In the first column we show scaling of the field level biases for each single-pair MV estimator relative to the MV $TT$ field-level bias. Here $p_\text{source}$ refers to the polarization fraction of the foreground, while $p_\text{CMB}\sim \sqrt{C^{EE}_\ell/C^{TT}_\ell}$, where $\ell$ should be evaluated where the lensing weights receive the majority of their support ($\ell\sim 3000$). In the second column we show the scaling for the field-level bias suppression when the MV estimators are substituted for LLS estimators. Note that there is no LLS $TB$ estimator to leading order in $L/\ell$, while the LLS $EB$ estimator is equivalent to the MV $EB$ estimator.
}
\label{tab:LLS}
\end{table}

We show the scaling of the field-level biases for all single-pair estimators in Table \ref{tab:LLS}. In the first column we show the scaling of the minimum variance (MV) field-level bias (relative to the MV $TT$ bias) with the polarization fractions of the foreground and CMB. In the second column we show the scaling of the field-level bias suppression for each LLS estimator. Note that while all LLS estimators reduce the bias by $\mathcal{O}((L/\ell)^2)$, the $TT$ and $TE$ estimators additionally reduce the bias by factors involving derivatives of the Fourier transform of the profile (e.g. $\ell u'_\ell \equiv \ell\, d u_\ell/d\ell$). For the special case of point sources ($u_\ell = 1$) these derivatives vanish, and the ($TT,\,TE$) LLS estimators are unbiased to all orders in $L/\ell$.

\begin{figure}[!h]
\centering
\includegraphics[width=\linewidth]{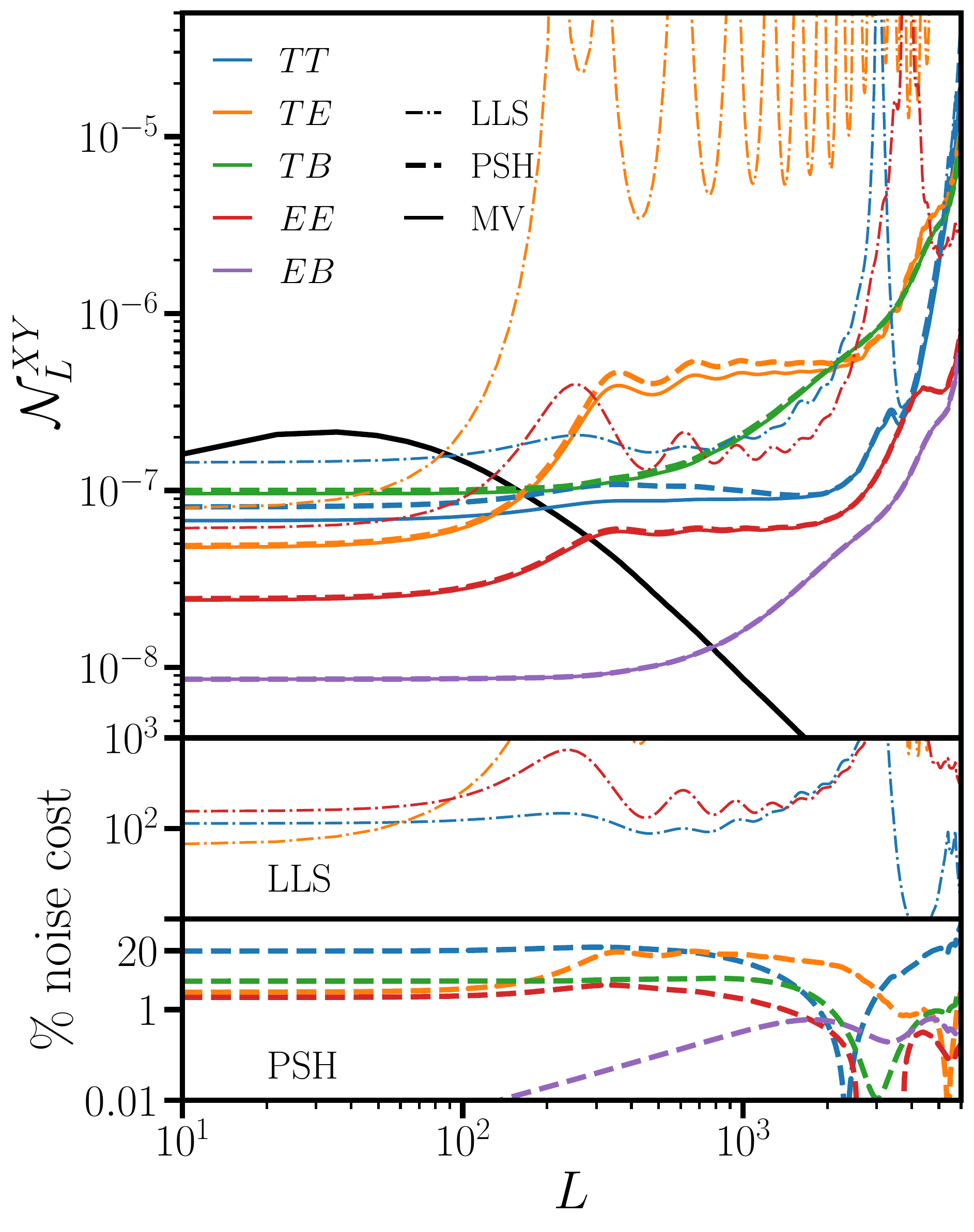}
\caption{The top panel shows the reconstruction noise (per Fourier mode) when reconstructing on a single pair of maps. The colors denote the pair $XY$. Solid, dashed and dashed-dotted curves represent minimum variance (MV), point source hardened (PSH) or large-lens suppressed (LLS) estimators respectively. $C^{\kappa\kappa}_L$ is shown in black. The bottom two panels show the percent noise cost for the LLS and PSH estimators relative to the appropriate MV estimator.
}
\label{fig:individual_noise_curves}
\end{figure}

\section{Noise properties}
\label{sec:noise}

Here we compute the variance of the estimators discussed in sections \ref{sec:theory}, \ref{sec:bias_hardening} and \ref{sec:LLS}. We consider a full-sky CMB experiment at a single frequency (150 GHz) with stage-4 like sensitivity ($\Delta_T = 1\,\,\mu$K-arcmin, $\Delta_P = \sqrt{2} \Delta_T$), and assume a Gaussian beam with a 1.4 arcmin FWHM. Throughout we take $\ell_{\text{max},T}=3500$ and $\ell_{\text{max},P} = 5000$.

In our total temperature power spectrum $\tilde{C}^{TT}_\ell$ we include the lensed CMB, instrument noise, and contributions from extragalactic foregrounds (tSZ, kSZ, CIB and radio point sources) using $\verb|LensQuEst|$\footnote{\href{https://github.com/EmmanuelSchaan/LensQuEst}{https://github.com/EmmanuelSchaan/LensQuEst}}, which adopts templates for the extragalactic foreground power spectra from \cite{2013JCAP...07..025D}. With the exception of tSZ $\times$ CIB, we neglect the cross-correlation of the components listed above. For polarization-only spectra and for cross-correlations of temperature and polarization, we include the lensed CMB and instrument noise (where appropriate) in our total power spectra, but neglect extragalactic foregrounds. We use $\verb|symlens|$\footnote{\href{https://github.com/simonsobs/symlens}{https://github.com/simonsobs/symlens}} to implement our estimators and to compute noise curves.

We first consider the variance of single-pair estimators, given by Eq.~\eqref{eq:noise_XY} and plotted in Fig.~\ref{fig:individual_noise_curves}. We see that point source hardening comes at a $1-20$\% noise cost (relative to the MV estimators) for most pairs of maps, and that bias hardening polarization-based estimators generically comes at a lower noise cost than the temperature-only case. In particular, the noise cost for hardening $EB$ against point sources is less than $0.1\%$ for all signal dominated scales ($L<1000$). By contrast, the single-pair LLS estimators come at more modest factor of $\sim 2$ noise cost for low $L$. As shown in Appendix \ref{sec:LLS_details}, all three non-trivial LLS estimators $(TT,TE,EE)$ are only sensitive to one of the two terms (monopole and quadrupole) in the lensing linear response in the large lens limit. Since these two terms are comparable in magnitude, the LLS estimators are only sensitive to roughly half of the lensing signal, providing some intuition for the approximate factor of $2$ cost in noise\footnote{That is, from Eq.~\eqref{eq:noise_large_lens} we see that the large-lens noise very roughly scales as (\# of multipoles)/(\# of multipoles)$^2$, assuming that each multipole is of a similar size (and sign).}.

\begin{figure}[!h]
\centering
\includegraphics[width=\linewidth]{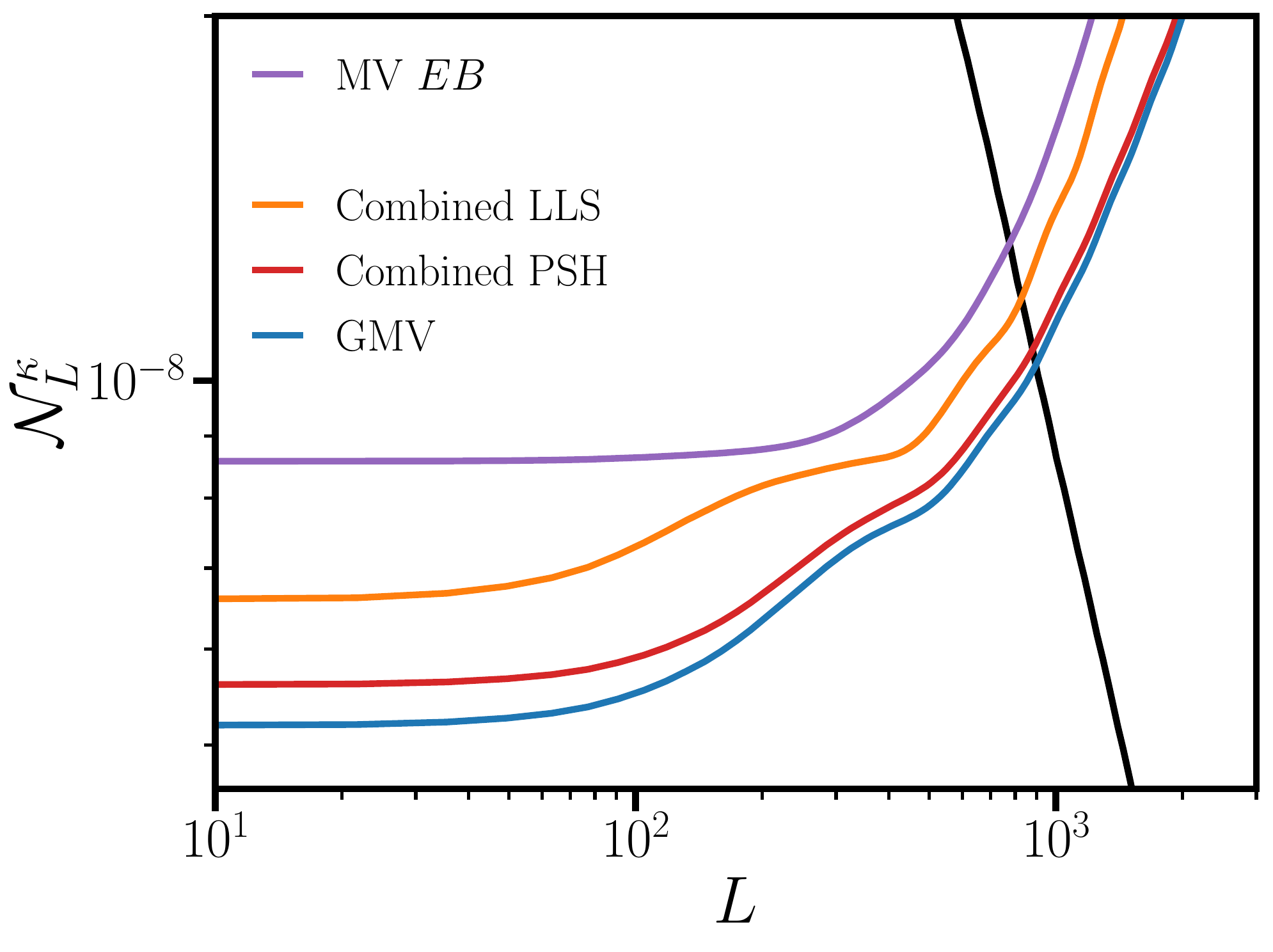}
\caption{
Lensing reconstruction noise for the optimal linear combination of single-pair LLS (orange) and point source hardened (PSH, red) estimators. The noise for the global MV (GMV) estimator is shown in blue. For reference we show the MV $EB$ noise (purple) and $C^{\kappa\kappa}$ (black). 
}
\label{fig:combined_noise}
\end{figure}

We next consider the reconstruction noise for combined estimators, shown in Fig.~\ref{fig:combined_noise}. In blue we show the reconstruction noise for the Global Minimum Variance estimator (i.e. Eq.~\eqref{eq:global_noise}), whose weights are explicitly given by Eq.~\eqref{eq:explicit_GMV}.
In orange and red we show the reconstruction noises for optimal linear combinations (see Appendix \ref{sec:combined}) of the single-pair LLS and point source hardened (PSH) estimators respectively. In particular, we find that near the transition of signal- to noise-domination ($L\sim 1000$) the combined PSH estimator's noise is only 4\% larger than the GMV's. The combined LLS estimator comes at more modest $\sim 20\%$ noise penalty.

In Appendix \ref{sec:combined} we explore the feasibility of constructing global forms of the LLS and PSH estimators, however, we find that an optimal linear combination of single-pair estimators is both more practical and nearly optimal in both cases.

\section{Bias reduction}
\label{sec:bias}

Biases to the CMB lensing power spectrum can be decomposed into primary, secondary and trispectrum components \cite{2018PhRvD..97b3512F, 2014JCAP...03..024O}. For simplicity consider the case of a single-pair estimator $\hat{\kappa}^{XY}_{\bm{L}}$ which is used to measure $\hat{C}^{\kappa\kappa}_L\propto \langle
\hat{\kappa}^{XY}_{\bm{L}} \hat{\kappa}^{XY}_{-\bm{L}}\rangle$ from a set of maps containing the lensed CMB $(\tilde{X},\tilde{Y})$ and the foregrounds of interest $(s^X,s^Y)$. Schematically, let $\hat{\kappa}[X,Y]$ denote reconstruction with the estimator $\hat{\kappa}^{XY}_{\bm{L}}$ on the maps $X_{\bm{\ell}}$, $Y_{\bm{L}-\bm{\ell}}$. Using this notation the primary, secondary and trispectrum biases take the form:
\beq
\bal
\label{eq:pst}
\text{Primary}  &= 2\big\langle \kappa\, \hat{\kappa}[s^X,s^Y]\big\rangle
\\
\text{Secondary}  &= 2\big\langle \hat{\kappa}[\tilde{X}_0,s^Y] \hat{\kappa}[\tilde{X}_1,s^Y]\big\rangle
\\&\,\,\,\,\,\,\,+
2\big\langle \hat{\kappa}[s^{X},\tilde{Y}_0] \hat{\kappa}[s^{X},\tilde{Y}_1]\big\rangle
\\&\,\,\,\,\,\,\,+
2\big\langle \hat{\kappa}[s^{X},\tilde{Y}_0] \hat{\kappa}[\tilde{X}_1,s^Y]\big\rangle
\\&\,\,\,\,\,\,\,+
2\big\langle \hat{\kappa}[s^{X},\tilde{Y}_1] \hat{\kappa}[\tilde{X}_0,s^Y]\big\rangle
\\
\text{Trispectrum}  &= \big\langle \hat{\kappa}[s^X,s^Y] \hat{\kappa}[s^X,s^Y]\big\rangle_c,
\eal
\eeq
where we have expanded $\tilde{X} = \tilde{X}_0 + \tilde{X}_1 +\cdots$ in powers of $\kappa$ to remove noise bias in our calculation of the secondary \cite{2019PhRvL.122r1301S}. The Gaussian component of the foreground four-point function is subtracted analytically to obtain the trispectrum, as indicated by the $c$ subscript.

\subsection{From field-level to power spectrum bias suppression}

Both bias hardened and LLS estimators aim to suppress the field-level response to foregrounds, which schematically take the form $ \int F g$, where $F$ is the lensing weights and $g$ is the foreground linear response. Here we relate these field-level responses to the primary and trispectrum biases to the CMB lensing power spectrum. 

The primary bias ultimately arises from a non-zero $\langle s_{\bm{\ell}} s_{\bm{L}-\bm{\ell}} \kappa_{-\bm{L}} \rangle$ bispectrum. Since it is trivially the case that $\langle s_{\bm{\ell}} s_{\bm{L}-\bm{\ell}} \kappa_{-\bm{L}} \rangle$ = $\langle \langle s_{\bm{\ell}} s_{\bm{L}-\bm{\ell}} \rangle' \kappa_{-\bm{L}} \rangle$, one can expand this bispectrum in terms of the linear foreground response and the foreground-lensing cross-power spectrum:
\beq
\label{eq:expanding_primary}
\langle 
s^X_{\bm{\ell}}
s^Y_{\bm{L}-\bm{\ell}}
\kappa_{-\bm{L}} 
\rangle
\propto 
\sum_Z
g^{XY,Z}_{\bm{\ell},\bm{L}-\bm{\ell}} 
\,\,
\langle s^Z_{\bm{L}} \kappa_{-\bm{L}} \rangle.
\eeq
As a consequence the primary bias is proportional to the field-level response. Thus choosing the weights $F$ to null the field-level response also nulls the primary bias.

Likewise, for a foreground obeying the assumptions discussed in Section \ref{sec:foreground_model}, we also find that the trispectrum can be expressed as a sum over power spectra, this time weighted by two factors of the foreground linear response:
\beq
\label{eq:expanding_trispectrum}
\langle 
s^X_{\bm{\ell}}
s^Y_{\bm{L}-\bm{\ell}}
s^M_{\bm{\ell}'}
s^N_{-\bm{L}-\bm{\ell}'}
\rangle 
\propto 
\sum_Z
\,
g^{XY,Z}_{\bm{\ell},\bm{L}-\bm{\ell}}
\,\,
g^{MN,Z}_{\bm{\ell}',-\bm{L}-\bm{\ell}'}
\,\,
\langle s^Z_{\bm{L}} s^Z_{-\bm{L}} \rangle,
\eeq
when at least one of the maps is a temperature map. Thus, for these cases, the trispectrum bias is proportional to the field-level response squared. Polarization-only trispectra also contain a term proportional to either $\cos(2 \Phi)$ or $\sin(2\Phi)$, where $\Phi \equiv \phi_{\bm{\ell}} + \phi_{\bm{L}-\bm{\ell}} - \phi_{\bm{\ell}'} - \phi_{-\bm{L}-\bm{\ell}'}$, in addition to those in Eq.~\eqref{eq:expanding_trispectrum}. In practice, we find that these extra terms are highly subdominant to those in Eq.~\eqref{eq:expanding_trispectrum}, as shown in Fig.~\ref{fig:biases}.

One can also obtain expressions analogous to Eq.~\eqref{eq:expanding_primary} for the secondary bias, which is sourced by the same $\langle ss\kappa\rangle$ bispectrum. However, the angular structure of the secondary bias differs significantly from that of the primary, and as a result there is no simple relation between the field-level response and the secondary bias. Thus we expect nulling the field-level bias to null both the primary and trispectrum, but not the secondary. We outline a new approach to null the secondary in Appendix \ref{sec:optimal_bias_hardening}, but leave further development to future work.

\begin{figure}[!h]
\centering
\includegraphics[width=\linewidth]{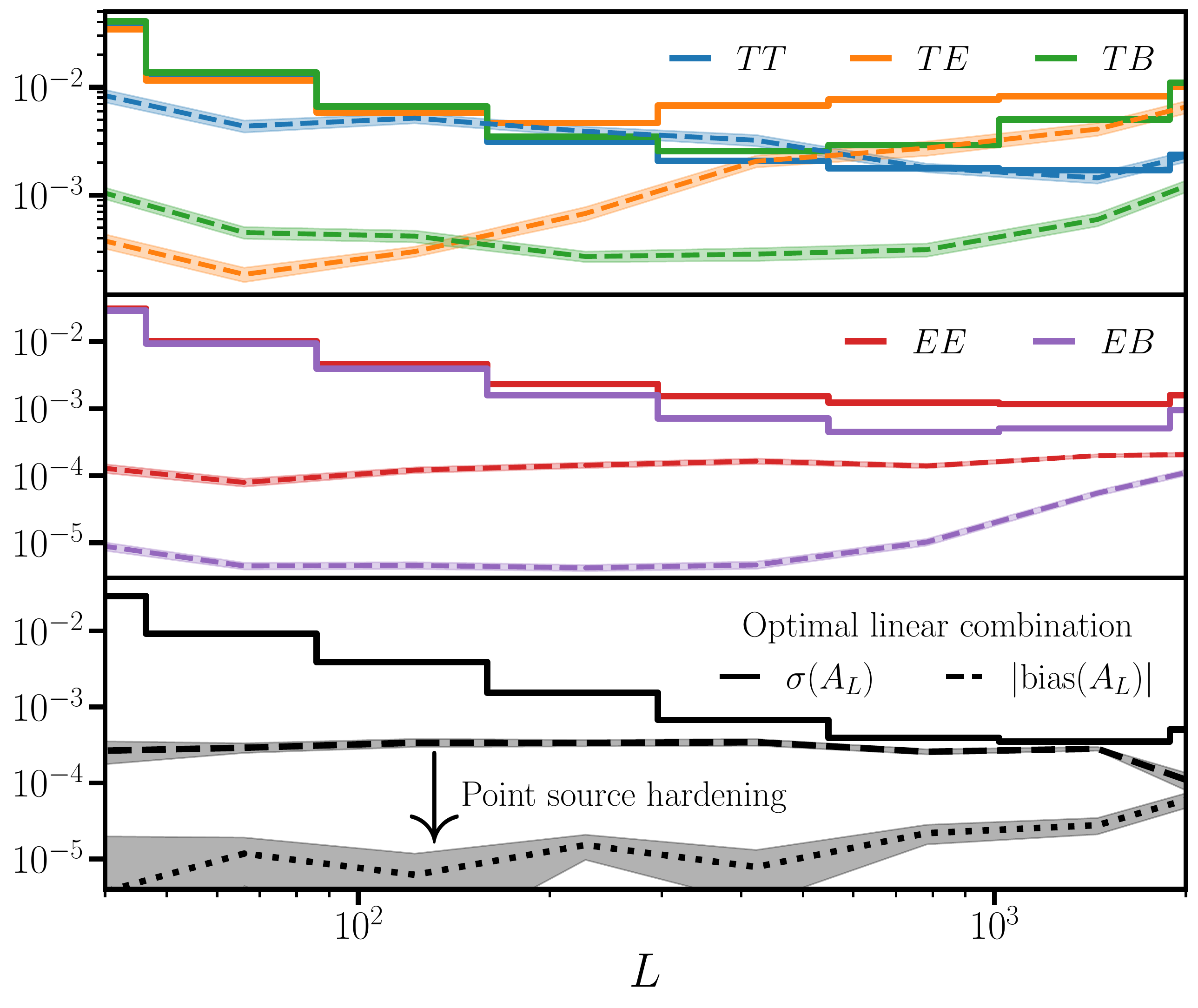}
\caption{\textit{Top two panels:} Solid lines show the $1\sigma$ error of the lensing amplitude ($A_L \equiv C^{\kappa\kappa,\text{measured}}_L/ C^{\kappa\kappa,\text{true}}_L$) for each single-pair minimum variance (MV) estimator. Colors denote pairs of temperature and polarization. Dashed lines show the simulated (absolute) biases to the MV estimators due to radio point sources, with the shaded regions denoting the $1\sigma$ scatter of the mean. \textit{Bottom:} Error (solid) and bias (dashed) to the lensing amplitude for the optimal linear combination of MV single-pair estimators. The dotted curve shows the bias for an optimal combination of point source hardened estimators. See sections \ref{sec:noise} and \ref{sec:bias} for details regarding the experimental configuration and simulations respectively. 
}
\label{fig:bias_and_noise}
\end{figure}

\begin{figure*}
\centering
\includegraphics[width=\linewidth]{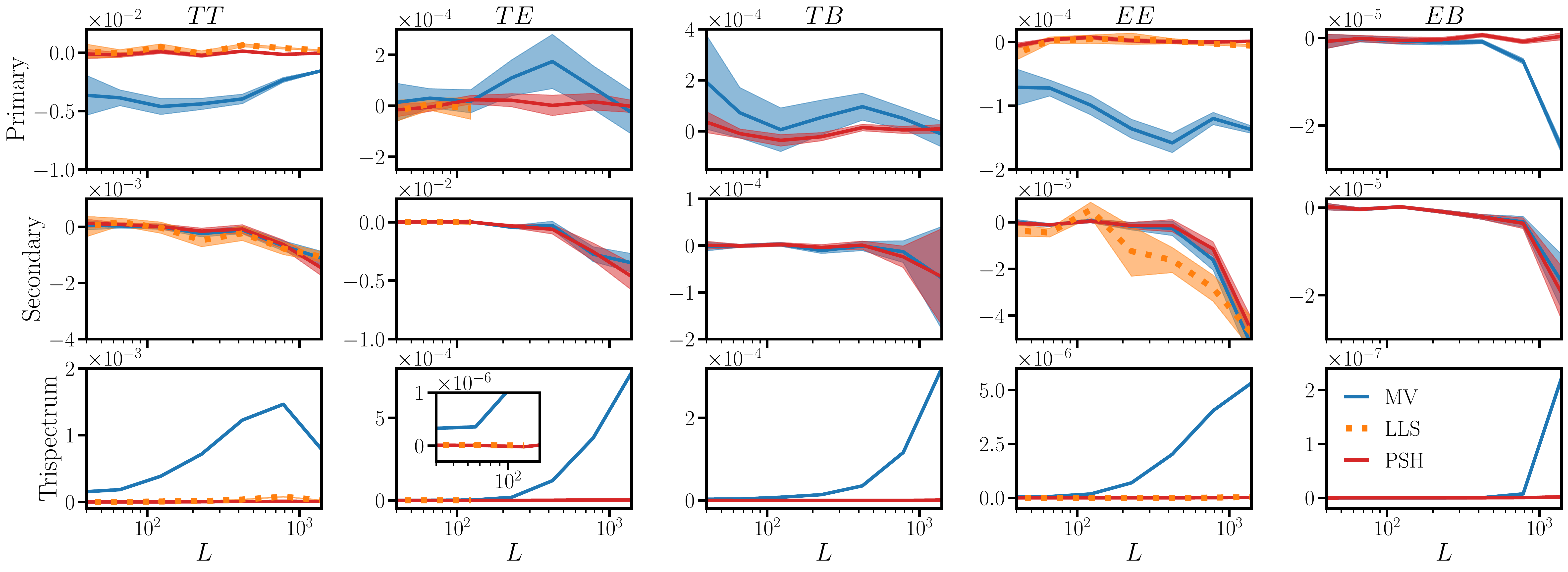}
\caption{Simulated relative biases to the CMB lensing power spectrum from point sources (assuming a 5 mJy flux cut and 3\% polarization fraction) for single-pair lensing estimators with $\ell_{\text{max},T} = 3500$ and $\ell_{\text{max},P}=5000$, and CMB-S4 like sensitivity. Columns correspond to different pairs of temperature and polarization, while the rows indicate primary, secondary or trispectrum biases. Biases for the Minimum Variance (MV), Large-Lens Suppressed (LLS) and Point Source Hardened (PSH) estimators are shown in blue, orange and red respectively. Note that for the $TE$ LLS estimator we only show biases for $L<120$ since this estimator's noise dramatically increases at smaller scales (see Fig.~\ref{fig:individual_noise_curves}), and that the $TE, TB$ primary biases vanish analytically. The assumed $L$-binning is depicted in Fig.~\ref{fig:bias_and_noise}, and the shaded regions denote the $1\sigma$ scatter of the mean biases across 40 cutouts. We stress that the $TT$ estimator here only shows the contribution from point sources: the overall bias is significantly larger when contributions from the CIB, tSZ and kSZ are included \cite{Sailer:2020lal}.
}
\label{fig:biases}
\end{figure*}

\subsection{Power spectrum biases: simulations}

To numerically estimate the biases in Eq.~\eqref{eq:pst} we use the WebSky simulations of both the lensing convergence \cite{Stein:2020its} and radio galaxies \cite{Li:2021ial} (in temperature). These simulations accurately reproduce number
counts measured from current CMB and radio telescopes, and have a realistic correlation between the galaxies and lensing convergence. Following \cite{2019PhRvL.122r1301S,Sailer:2020lal} we mask sources brighter than 5 mJy by fulling excluding the pixels associated with those sources. 

We generate mock polarization data by assigning random polarization angles (sampled from a uniform distribution) to each source. For simplicity we assume that all sources have the same $3\%$ polarization fraction. The measured biases presented below tend to scale simply with moments of the polarization fraction distribution, e.g. the $EE$ trispectrum bias $\propto\langle p^4\rangle$. Thus one can scale our results to any desired polarization fraction distribution.  

We divide a subsample of the WebSky simulation into 40 square 15 degree cutouts using $\verb|pixell|$\footnote{\href{https://github.com/simonsobs/pixell}{https://github.com/simonsobs/pixell}}. For each cutout we generate mock Gaussian CMB maps with $\verb|LensQuEst|$ and lens these maps using the corresponding WebSky convergence cutout. We run the lensing reconstruction for each single-pair estimator under the flat sky approximation using $\verb|symlens|$, assuming the S4-like configuration described in section \ref{sec:noise}. 

First, in top two panels of Fig.~\ref{fig:bias_and_noise} we show the sensitivity\footnote{The sensitivities quoted here are for the assumed $L-$binning in Fig.~\ref{fig:bias_and_noise}, and would be more significant if broader bins were used.} to the lensing amplitude of each single-pair minimum variance (MV) estimator, along with the total bias (primary + secondary + trispectrum) from polarized point sources. We note that these biases are significant ($\sim 0.5-1\sigma$) for both the $TT$ and $TE$ estimators, while for the remaining single-pair estimators they can approach $\sim0.1\sigma$ at small $(L>500)$ scales. The bias to the optimal linear combination of single-pair MV estimators (bottom panel) is $\sim 1\sigma$ for $L>500$, and non-negligible on larger scales. These biases are suppressed by up to two orders of magnitude when one instead takes an optimal linear combination of point source hardened estimators, ensuring $\text{bias} < \sigma/50$ for $L<500$ and $< \sigma/20$ for $500 < L < 1000$. We show an analogous figure for a Simons Observatory-like survey in Appendix \ref{sec:SO_biases}, and again find that the biases to $TT$ and $TE$ are significant ($\sim 1\sigma$), while for the remaining estimators the biases are negligible ($<\sigma/100$).

In Fig.~\ref{fig:biases} we show the individual (relative) primary, secondary and trispectrum biases for the Minimum Variance (MV; blue), Point Source Hardened (PSH; orange), and Large-Lens Suppressed estimators (LLS; red). We find that both point source hardening and large-lens suppression effectively null the primary and trispectrum biases for all single-pair estimators, but have little impact on the secondary bias, as expected. Since the total $TT,\,TB,\,EE,$ and $EB$ biases receive sizeable contributions from the primary and trispectrum terms, one significantly reduces the overall bias for these estimators. However, the $TE$ bias is dominated by the secondary $\langle s^T s^T \kappa\rangle$ contribution, and thus the $TE$ bias is only marginally reduced using these techniques. If necessary one can null this secondary contribution using the technique outlined in Appendix \ref{sec:optimal_bias_hardening}. As indicated in the bottom panel of Fig.~\ref{fig:bias_and_noise}, the $TE$ secondary is a subdominant contribution to the bias from the optimal combination of PSH estimators, and we therefore expect that an optimal linear combination of PSH estimators is sufficient for a robust and precise measurement of $C^{\kappa\kappa}$ with future surveys.

We note that there is another population of ``Infrared'' (IR) point sources which are dominant at higher frequency and have not been included in the simulations described here, since their polarization fraction is expected to be smaller than for radio sources \cite{Datta:2018oae, SPT:2019wyt}. The same techniques discussed here should also mitigate the impact of polarized IR sources with similar effectiveness.

\section{Discussion and Conclusions}
\label{sec:conclusions}

In this paper we generalized previous work to polarization, showing that geometric methods are effective at reducing the bias to CMB lensing in both temperature and polarization, even on single-frequency maps. We used state-of-the art simulations of radio point sources to estimate the size of these biases and the effectiveness of our mitigation strategies. We have found that the point-source bias in polarization is relatively small for upcoming experiments such as Simons Observatory (Appendix \ref{sec:SO_biases}), but it can become potentially significant on small scales for future low-noise experiments (Fig.~\ref{fig:bias_and_noise}). We have shown that point source hardened estimators, which come at a negligible noise cost, can very effectively control these biases for all experiments in the foreseeable future. We find that LLS estimators are equally as effective at reducing these biases, however, generically come at a significantly higher noise cost.

We note that patchy reionization is also expected to give a contribution to the temperature fluctuations (through the kSZ effect) and to polarization, through patchy screening and scattering \cite{Dvorkin:2008tf}. While we expect the biases induced by reionization to be subdominant (this has been shown explicitly for temperature in \cite{Cai:2021hnb}), we note that that the majority of the reionization bubbles will be sub-beam in size, and therefore we expect PSH to decrease any possible bias. A full study of this is left to future work.

In this study we have not explored the impact of \textit{galactic} foregrounds and potential mitigation strategies. In \cite{Beck:2020dhe}, the authors show that foreground-induced biases can be significant in polarization, and consider multi-frequency cleaning as a partial mitigation strategy. While we note that the simple geometry of point sources allowed us to analytically write estimators with zero response to them, some of the same techniques presented here may prove useful when dealing with galactic foregrounds. The lensing-induced distortion provides a consistency between the different angular multipoles discussed in Section \ref{sec:LLS}. Due to their complex geometry, galactic foregrounds are unlikely to match such a consistency and therefore they may be distinguishable using geometric methods such as the ones discussed in this paper. Finally, it's worth noting that the same dust seen in emission in the CMB maps also causes extinction at optical wavelengths, potentially modulating the observed galaxy field and creating biases to cross-correlations. This further motivates this line of research, which is left to a future investigation.

\section*{Acknowledgments}

We thank Julien Carron, Antón Baleato Lizancos, Zack Li, Marius Millea, Toshiya Namikawa and Giuseppe Puglisi for useful discussions. 
NS and SF are supported by the Physics Division of Lawrence Berkeley National Laboratory and by the U.S. Department of Energy (DOE), Office of Science, under contract No.
DE-AC02-05CH11231.
ES was supported by the Chamberlain fellowship at Lawrence Berkeley National
Laboratory, and is now supported by SLAC National Accelerator Laboratory.

\bibliographystyle{prsty.bst}
\bibliography{main}

\onecolumngrid
\appendix

\section{Expansions in the large-lens limit}
\label{sec:taylor_expansions}

Any function of $|\bm{L}-\bm{\ell}|$ can be Taylor expanded in powers of $L/\ell$ to give:
\beq
\label{eq:generic_taylor}
u_{|\bm{L}-\bm{\ell}|}
 = u_\ell - L u'_\ell \cos(\theta_{\bm{L},\bm{\ell}})   
 + \frac{1}{4}\left(\frac{L}{\ell}\right)^2
 \bigg[ D^+_\ell u_\ell + \cos(2\theta_{\bm{L},\bm{\ell}})D^-_\ell u_\ell\bigg]
\eeq
with corrections going as $\mathcal{O}((L/\ell)^3)$, where in Eq.~\eqref{eq:generic_taylor} we have defined the operators
\beq
D^\pm_\ell \equiv \ell^2 \frac{d^2}{d\ell^2} \pm \frac{d}{d\ln\ell}.
\eeq
We can similarly Taylor expand the following geometric factors to quadratic order:
\beq
\bal
\cos(2\theta_{\bm{\ell},\bm{L}-\bm{\ell}}) &= 1- \left(\frac{L}{\ell}\right)^2 + \left(\frac{L}{\ell}\right)^2 \cos(2\theta_{\bm{L},\bm{\ell}})
\\
\sin(2\theta_{\bm{\ell},\bm{L}-\bm{\ell}}) &= 2 \frac{L}{\ell} \sin(\theta_{\bm{L},\bm{\ell}} )+\left(\frac{L}{\ell}\right)^2 \sin(2 \theta_{\bm{L},\bm{\ell}} )
\\
\cos(2\theta_{\bm{L},\bm{L}-\bm{\ell}}) &= 
\cos(2\theta_{\bm{L},\bm{\ell}}) 
+
\frac{L}{\ell}
\bigg[
\cos(3\theta_{\bm{L},\bm{\ell}})
-
\cos(\theta_{\bm{L},\bm{\ell}})
\bigg]
+
\left(\frac{L}{\ell}\right)^2
\bigg[
\cos(4\theta_{\bm{L},\bm{\ell}})
-
\cos(2\theta_{\bm{L},\bm{\ell}})
\bigg]
\\
\sin(2\theta_{\bm{L},\bm{L}-\bm{\ell}}) &= 
\sin(2\theta_{\bm{L},\bm{\ell}}) 
+
\frac{L}{\ell}
\bigg[
\sin(3\theta_{\bm{L},\bm{\ell}})
-
\sin(\theta_{\bm{L},\bm{\ell}})
\bigg]
+
\left(\frac{L}{\ell}\right)^2
\bigg[
\sin(4\theta_{\bm{L},\bm{\ell}})
-
\sin(2\theta_{\bm{L},\bm{\ell}})
\bigg]
\eal
\eeq

Using these expressions one can Taylor expand the linear responses in Eq.~\eqref{eq:responses}:
\beq
\bal
\label{eq:taylor_expanded_responses}
f^{TT}_{\bm{\ell},\bm{L}-\bm{\ell}}
&= 
C^{TT}_\ell
\left[
\textcolor{matplotlib_red}{
\frac{d\ln \ell^2 C^{TT}_\ell}{d\ln\ell}
}
+
\textcolor{matplotlib_blue}{
\cos(2\theta_{\bm{L},\bm{\ell}})
\frac{d\ln C^{TT}_\ell}{d\ln\ell}
}
\right]
-
\frac{1}{2}
\frac{L}{\ell}
\left[
\textcolor{matplotlib_red}{
D^+_\ell C^{TT}_\ell
}
+
\textcolor{matplotlib_blue}{
4\frac{d C^{TT}_\ell}{d\ln\ell} \cos(\theta_{\bm{L},\bm{\ell}})
}
+
\textcolor{matplotlib_blue}{
\cos(2 \theta_{\bm{L},\bm{\ell}})
D^-_\ell C^{TT}_\ell
}
\right]
\\
\\
f^{TE}_{\bm{\ell},\bm{L}-\bm{\ell}}
&= 
C^{TE}_\ell
\left[
\textcolor{matplotlib_blue}{
\frac{d\ln \ell^2 |C^{TE}_\ell|}{d\ln\ell}
}
+
\textcolor{matplotlib_red}{
\cos(2\theta_{\bm{L},\bm{\ell}})
\frac{d\ln C^{TE}_\ell}{d\ln\ell}
}
\right]
-
\frac{1}{2}
\frac{L}{\ell}
\bigg[
\textcolor{matplotlib_blue}{
\big(D^+_\ell+4\big) C^{TE}_\ell
}
+
\textcolor{matplotlib_blue}{
4\frac{d C^{TE}_\ell}{d\ln\ell} \cos(\theta_{\bm{L},\bm{\ell}})
}
+
\textcolor{matplotlib_red}{
\cos(2 \theta_{\bm{L},\bm{\ell}})
\big(D^-_\ell -4\big) C^{TE}_\ell
}
\bigg]
\\
\\
f^{TB}_{\bm{\ell},\bm{L}-\bm{\ell}}
&=
\textcolor{matplotlib_red}{
2\sin(2\theta_{\bm{L},\bm{\ell}})C^{TE}_\ell
}
+
\textcolor{matplotlib_red}{
\frac{L}{\ell} C^{TE}_\ell 
\bigg[
\sin(\theta_{\bm{L},\bm{\ell}}) 
+ 
\sin(3\theta_{\bm{L},\bm{\ell}})
\bigg]
}
\\
\\
f^{EE}_{\bm{\ell},\bm{L}-\bm{\ell}}
&= 
C^{EE}_\ell
\left[
\textcolor{matplotlib_red}{
\frac{d\ln \ell^2 C^{EE}_\ell}{d\ln\ell}
}
+
\textcolor{matplotlib_blue}{
\cos(2\theta_{\bm{L},\bm{\ell}})
\frac{d\ln C^{EE}_\ell}{d\ln\ell}
}
\right]
-
\frac{1}{2}
\frac{L}{\ell}
\left[
\textcolor{matplotlib_red}{
D^+_\ell C^{EE}_\ell
}
+
\textcolor{matplotlib_blue}{
4\frac{d C^{EE}_\ell}{d\ln\ell} \cos(\theta_{\bm{L},\bm{\ell}})
}
+
\textcolor{matplotlib_blue}{
\cos(2 \theta_{\bm{L},\bm{\ell}})
D^-_\ell C^{EE}_\ell
}
\right]
\\
\\
f^{EB}_{\bm{\ell},\bm{L}-\bm{\ell}}
&=
\textcolor{matplotlib_blue}{
2\sin(2\theta_{\bm{L},\bm{\ell}})C^{EE}_\ell
}
+
\frac{L}{\ell} C^{EE}_\ell 
\bigg[
\textcolor{matplotlib_blue}{
\sin(\theta_{\bm{L},\bm{\ell}}) 
}
+ 
\textcolor{matplotlib_blue}{
\sin(3\theta_{\bm{L},\bm{\ell}}),
}
\bigg]
\eal
\eeq
with corrections going as $\mathcal{O}((L/\ell)^2)$.
The colors here denote terms which are orthogonal to the relevant foreground linear response functions to $\mathcal{O}((L/\ell)^2)$ (blue), and those which aren't (red). That is, within the LLS approach we wish to retain sensitivity to the (clean) blue terms, while killing sensitivity to the red.

Likewise, one can Taylor expand the angular dependent pieces of the foreground linear responses (Table \ref{tab:response}) to give:
\beq
\bal
\label{eq:taylor_expanded_foreground_responses}
g^{TT,T}_{\bm{\ell},\bm{L}-\bm{\ell}} 
&= \mathcal{A}  \frac{u_\ell}{u_L}
\bigg\{
u_\ell - L u'_\ell \cos(\theta_{\bm{L},\bm{\ell}})
+\frac{1}{4}\left(\frac{L}{\ell}\right)^2
\bigg[ D^+_\ell  + \cos(2\theta_{\bm{L},\bm{\ell}})D^-_\ell \bigg]u_\ell
\bigg\}
\\
g^{TE,E}_{\bm{\ell},\bm{L}-\bm{\ell}} &= \mathcal{B} \frac{u_\ell}{u_L} 
\bigg\{
u_\ell \cos(2\theta_{\bm{L},\bm{\ell}})
+
\frac{L}{\ell}
\bigg[
(u_\ell - \ell u'_\ell/2)
\cos(3\theta_{\bm{L},\bm{\ell}})
-
(u_\ell + \ell u'_\ell/2)
\cos(\theta_{\bm{L},\bm{\ell}})
\bigg]\\
&\,\,\,\,\,\,\,\,\,\,\,\,\,\,\,\,\,\,\,\,\,\,\,\,\,\,\,\,+
\frac{1}{4}\left(\frac{L}{\ell}\right)^2
\bigg[
D^-_\ell u_\ell/2+
\cos(2\theta_{\bm{L},\bm{\ell}}) (D^+_\ell u_\ell - 4u_\ell) 
+
(4 u_\ell+ D^-_\ell u_\ell/2 ) \cos(4\theta_{\bm{L},\bm{\ell}})
\bigg]
\bigg\}
\\
g^{TE,B}_{\bm{\ell},\bm{L}-\bm{\ell}} &= -\mathcal{B}   \frac{u_\ell}{u_L} 
\bigg\{
u_\ell \sin(2\theta_{\bm{L},\bm{\ell}}) +\frac{L}{\ell} \bigg[
(u_\ell - \ell u'_\ell/2) \sin(3\theta_{\bm{L},\bm{\ell}})
- (u_\ell + \ell u'_\ell/2) \sin(\theta_{\bm{L},\bm{\ell}})
\bigg]
\\
&\,\,\,\,\,\,\,\,\,\,\,\,\,\,\,\,\,\,\,\,\,\,\,\,\,\,\,\,+
\frac{1}{4}\left(\frac{L}{\ell}\right)^2
\bigg[
\sin(2\theta_{\bm{L},\bm{\ell}}) (D^+_\ell u_\ell - 4u_\ell) + \sin(4\theta_{\bm{L},\bm{\ell}}) (4u_\ell + D^-_\ell u_\ell/2)
\bigg]
\bigg\}
\\
g^{EE,T}_{\bm{\ell},\bm{L}-\bm{\ell}} &= \mathcal{C}  \frac{u_\ell}{u_L}
\bigg\{
u_\ell - L u'_\ell \cos(\theta_{\bm{L},\bm{\ell}}) 
+\frac{1}{4}\left(\frac{L}{\ell}\right)^2
\bigg[ (D^+_\ell-4)u_\ell + \cos(2\theta_{\bm{L},\bm{\ell}})(D^-_\ell +4)u_\ell\bigg]
\bigg\}
\\
g^{EB,T}_{\bm{\ell},\bm{L}-\bm{\ell}} &= \mathcal{C}  \frac{u_\ell}{u_L}
\bigg\{
2 \frac{L}{\ell} u_\ell \sin(\theta_{\bm{L},\bm{\ell}})
- 
\left(\frac{L}{\ell}\right)^2 
\bigg[
\ell u'_\ell 
+
 u_\ell 
\bigg]
\sin(2 \theta_{\bm{L},\bm{\ell}} ) 
\bigg\},
\eal
\eeq
while the remaining two responses satisfy
$g^{TB,E}_{\bm{\ell},\bm{L}-\bm{\ell}}
=
-g^{TE,B}_{\bm{\ell},\bm{L}-\bm{\ell}}$
and 
$
g^{TB,B}_{\bm{\ell},\bm{L}-\bm{\ell}} 
=
g^{TE,E}_{\bm{\ell},\bm{L}-\bm{\ell}}. 
$
Terms in $\{\cdots\}$'s are accurate to $\mathcal{O}((L/\ell)^3)$.

\section{Explicit expressions for the global minimum variance weights}
\label{sec:GMV_explicit}

Let $\bm{F}^\text{GMV}$ denote the minimum variance set of weights (Eq.~\eqref{eq:MV_weights}) for a quadratic estimator taking the form of Eq.~\eqref{eq:general_QE}, where the index $i$ runs over $T,E,B$ at a single frequency:
\beq
\bm{F}^\text{GMV}_{\bm{\ell},\bm{L}-\bm{\ell}} = 
\begin{pmatrix}
F^{\text{GMV},TT}_{\bm{\ell},\bm{L}-\bm{\ell}} & F^{\text{GMV},TE}_{\bm{\ell},\bm{L}-\bm{\ell}} & F^{\text{GMV},TB}_{\bm{\ell},\bm{L}-\bm{\ell}} \\
F^{\text{GMV},TE}_{\bm{L}-\bm{\ell},\bm{\ell}} & F^{\text{GMV},EE}_{\bm{\ell},\bm{L}-\bm{\ell}} & F^{\text{GMV},EB}_{\bm{\ell},\bm{L}-\bm{\ell}} \\
F^{\text{GMV},TB}_{\bm{L}-\bm{\ell},\bm{\ell}} & F^{\text{GMV},EB}_{\bm{L}-\bm{\ell},\bm{\ell}} & 0 
\end{pmatrix}.
\eeq
Assuming $\tilde{C}^{TB} = \tilde{C}^{EB} = 0$, we find:
\beq
\bal
\label{eq:explicit_GMV}
F^{\text{GMV},TT}_{\bm{\ell},\bm{L}-\bm{\ell}}
&=
\left[
\tilde{C}^{EE}_\ell \tilde{C}^{EE}_{|\bm{L}-\bm{\ell}|}
f^{TT}_{\bm{\ell},\bm{L}-\bm{\ell}}
+
\tilde{C}^{TE}_\ell \tilde{C}^{TE}_{|\bm{L}-\bm{\ell}|}
f^{EE}_{\bm{\ell},\bm{L}-\bm{\ell}}
-
\tilde{C}^{EE}_\ell \tilde{C}^{TE}_{|\bm{L}-\bm{\ell}|}
f^{TE}_{\bm{\ell},\bm{L}-\bm{\ell}}
-
\tilde{C}^{TE}_\ell \tilde{C}^{EE}_{|\bm{L}-\bm{\ell}|}
f^{TE}_{\bm{L}-\bm{\ell},\bm{\ell}}
\right]
\frac{1}{
2 \mathcal{D}_\ell \mathcal{D}_{|\bm{L}-\bm{\ell}|}
}
\\
F^{\text{GMV},TE}_{\bm{\ell},\bm{L}-\bm{\ell}}
&=
\left[
\tilde{C}^{EE}_\ell \tilde{C}^{TT}_{|\bm{L}-\bm{\ell}|}
f^{TE}_{\bm{\ell},\bm{L}-\bm{\ell}}
+
\tilde{C}^{TE}_\ell \tilde{C}^{TE}_{|\bm{L}-\bm{\ell}|}
f^{TE}_{\bm{L}-\bm{\ell},\bm{\ell}}
-
\tilde{C}^{TE}_\ell \tilde{C}^{TT}_{|\bm{L}-\bm{\ell}|}
f^{EE}_{\bm{\ell},\bm{L}-\bm{\ell}}
-
\tilde{C}^{EE}_\ell \tilde{C}^{TE}_{|\bm{L}-\bm{\ell}|}
f^{TT}_{\bm{\ell},\bm{L}-\bm{\ell}}
\right]
\frac{1}{
2 \mathcal{D}_\ell \mathcal{D}_{|\bm{L}-\bm{\ell}|}
}
\\
F^{\text{GMV},TB}_{\bm{\ell},\bm{L}-\bm{\ell}}
&=
\left[
\tilde{C}^{EE}_\ell 
f^{TB}_{\bm{\ell},\bm{L}-\bm{\ell}}
-
\tilde{C}^{TE}_\ell 
f^{EB}_{\bm{\ell},\bm{L}-\bm{\ell}}
\right]
\frac{1}{
2 \mathcal{D}_\ell
\tilde{C}^{BB}_{|\bm{L}-\bm{\ell}|}
}
\\
F^{\text{GMV},EE}_{\bm{\ell},\bm{L}-\bm{\ell}}
&=
\left[
\tilde{C}^{TE}_\ell \tilde{C}^{TE}_{|\bm{L}-\bm{\ell}|}
f^{TT}_{\bm{\ell},\bm{L}-\bm{\ell}}
+
\tilde{C}^{TT}_\ell \tilde{C}^{TT}_{|\bm{L}-\bm{\ell}|}
f^{EE}_{\bm{\ell},\bm{L}-\bm{\ell}}
-
\tilde{C}^{TE}_\ell \tilde{C}^{TT}_{|\bm{L}-\bm{\ell}|}
f^{TE}_{\bm{\ell},\bm{L}-\bm{\ell}}
-
\tilde{C}^{TT}_\ell \tilde{C}^{TE}_{|\bm{L}-\bm{\ell}|}
f^{TE}_{\bm{L}-\bm{\ell},\bm{\ell}}
\right]
\frac{1}{2 \mathcal{D}_\ell \mathcal{D}_{|\bm{L}-\bm{\ell}|}}
\\
F^{\text{GMV},EB}_{\bm{\ell},\bm{L}-\bm{\ell}}
&=
\left[
\tilde{C}^{TT}_\ell 
f^{EB}_{\bm{\ell},\bm{L}-\bm{\ell}}
-
\tilde{C}^{TE}_\ell 
f^{TB}_{\bm{\ell},\bm{L}-\bm{\ell}}
\right]
\frac{1}{
2 \mathcal{D}_\ell
\tilde{C}^{BB}_{|\bm{L}-\bm{\ell}|}
}
\eal
\eeq
where we have defined $\mathcal{D}_\ell \equiv\tilde{C}^{TT}_\ell \tilde{C}^{EE}_\ell - (\tilde{C}^{TE}_\ell)^2$. We note that our convention of summing over both permutations of pairs, e.g. both $TE$ and $ET$, differs from that used in \cite{Maniyar:2021msb}. As a result our $TE$, $TB$ and $EB$ weights differ by a conventional factor of $2$. Additionally we find that the last term of $F^{\text{GMV},TE}$ differs from that in \cite{Maniyar:2021msb}, where the arguments $\bm{\ell}$ and $\bm{L}-\bm{\ell}$ have been flipped.

\section{Cleaning combined estimators}
\label{sec:combined}

Here we explore the feasibility of cleaning the GMV estimator via bias hardening and large-scale suppression. We find that for both cases one should in practice take an optimal linear combination of the relevant single-pair estimators, rather than to harden (or suppress) a GMV estimator.

\subsection{Bias hardening}
\label{sec:GPSH}

In section \ref{sec:wrong_profile} we found that the optimal temperature-only bias hardened estimator is insensitive to the amplitude $\mathcal{A}$ of the foregrounds. Here we show that this is always be the case when bias hardening a single-pair lensing estimator.

\textit{Proof:} In general, the optimal bias hardened weights for an estimator constructed from a single pair $X,Y$ takes the (schematic) form:
\beq
\label{eq:BH_XY_weights}
F^{XY} + \alpha^T\, G^{XY,T} + \alpha^E\, G^{XY,E} + \alpha^B\, G^{XY,B},
\eeq
where $F^{XY}$ is the MV lensing weight and $G^{XY,Z}$ is the MV weight for the foreground in the $Z$'th map. WLOG we will assume that the weights $G^{XY,Z}\propto g^{XY,Z}$ have been rescaled to remove the unknown prefactor ($\mathcal{A}$, $\mathcal{B}$ or $\mathcal{C}$) in the foreground response $g$, which can be thought of as being absorbed into coefficients $(\alpha^T,\alpha^E,\alpha^B)$. These coefficients are chosen so that the bias hardened estimator has zero response to $s^T$, $s^E$ and $s^B$. This is enforced by integrating Eq.~\eqref{eq:BH_XY_weights} against $g^{XY,T}$, and forcing this integral to vanish (and similarly for $g^{XY,E}$ and $g^{XY,B}$). Since these integral constraints are set to zero, we are always free to divide out the unknown prefactor introduced by the foreground response $g^{XY,Z}$ in all three constraint equations. We can then solve for the coefficients $(\alpha^T,\alpha^E,\alpha^B)$ without any knowledge of $(\mathcal{A},\mathcal{B},\mathcal{C})$. Thus Eq.~\eqref{eq:BH_XY_weights} is independent of $(\mathcal{A},\mathcal{B},\mathcal{C})$. $\square$

This is no longer true, however, when bias hardening the GMV estimator. 
In this case, the optimal bias hardened weights take the same form as Eq.~\eqref{eq:BH_XY_weights}, but with $F^{XY}$ replaced by the GMV lensing weights, and similarly for the foreground weights (e.g. the GMV weights for $s^T$ takes the same form as Eq.~\eqref{eq:MV_weights}, but with $f^{ij}$ replaced by $g^{XY,T}$, see Appendix \ref{sec:optimal_bias_hardening}). From Table~\ref{tab:response} note that the GMV weights for $s^E$ and $s^B$ are proportional to $\mathcal{B}$. However, the GMV weights for $s^T$ contain a linear combination of $\mathcal{A}$ and $\mathcal{C}$. Consequentially, one cannot bias harden the GMV $\kappa$ estimator without assuming the ratio $\mathcal{C}/\mathcal{A} \simeq \langle p_i^2\rangle/2$. 

To avoid assuming this ratio, one can simply take an optimal linear combination of the single-pair bias hardened estimators, which takes the form $\sum_i w^i_{\bm{L}} \hat{\kappa}^{\text{BH},i}_{\bm{L}}$, where $i\in \{TT,TE,TB,EE,EB\}$ and $w_i = \sum_j C^{-1}_{ji} / \sum_{ij} C^{-1}_{ij}$, where $C_{ij}$ is the covariance between $\hat{\kappa}^{\text{BH},i}$ and $\hat{\kappa}^{\text{BH},j}$:
\beq
C^{XY,MN}_{\bm{L}} = N^{XY}_{\bm{L}} N^{MN}_{\bm{L}} \int_{\bm{\ell}} F^{XY}_{\bm{\ell},\bm{L}-\bm{\ell}}\bigg[F^{MN}_{\bm{\ell},\bm{L}-\bm{\ell}} \tilde{C}^{XM}_\ell \tilde{C}^{YN}_{|\bm{L}-\bm{\ell}|} + F^{MN}_{\bm{L}-\bm{\ell},\bm{\ell}} \tilde{C}^{XN}_\ell \tilde{C}^{YM}_{|\bm{L}-\bm{\ell}|}\bigg].
\eeq
The noise of this optimal linear combination is given by $1/\sum_{ij}C^{-1}_{ij}$.

This simple linear combination of bias hardened estimators is very close to optimal (while remaining practical) for the following reason. The GMV estimator can be separated into two statistically independent GMV estimators constructed from $\{TT,TE,EE\}$ and $\{TB,EB\}$ \cite{Maniyar:2021msb}. Following the same argument as above one cannot bias harden the former without assuming the ratio $\mathcal{C}/\mathcal{A}$. To get around this, one can further decompose the set into $\{TT,TE\}$ and $\{EE\}$. In practice the GMV estimator constructed from $\{TT,TE\}$ is not FFT-able. To make this estimator FFT-able one can neglect the cross-correlation $\tilde{C}^{TE}$, however, in this approximation the $\{TT,TE\}$ GMV estimator reduces to the optimal linear combination of $\hat{\kappa}^{TT}$ and $\hat{\kappa}^{EE}$. Thus, to bias harden the $\{TT,TE,EE\}$ GMV estimator in an FFT-able way without assuming $\mathcal{C}/\mathcal{A}$ one must fully decompose $\{TT,TE,EE\}$ into an optimal linear combination of $TT$, $TE$ and $EE$. 

While one can in principle bias harden the GMV estimator constructed from $\{TB,EB\}$ in practice the gains for doing so (as opposed to an optimal linear combination) are expected to be minimal. We have checked that the reconstruction noise for the GMV constructed from $\{TB,EB\}$ and the optimal linear combination of $TB$ and $EB$ differ by less than half a percent for $L<5000$.

\subsection{Large-lens suppression}

One can generalize the single-pair LLS estimator to a global form in a straightforward way. However, as we show below, we find that an optimal linear combination of single-pair LLS estimators is practically more useful. 

From Eqs.~\eqref{eq:LLS_temp} and \eqref{eq:LLS_wpol} a natural ansatz for the global LLS weights is:
\beq
\bm{F}^\text{LLS}_{\bm{\ell},\bm{L}-\bm{\ell}}
=
\begin{pmatrix}
A_{L,\ell}\, \cos(2\theta_{\bm{L},\bm{\ell}}) & B_{L,\ell} & 0\\
B_{L,|\bm{L}-\bm{\ell}|} & C_{L,\ell}\, \cos(2\theta_{\bm{L},\bm{\ell}}) & F^{\text{MV},EB}_{\bm{\ell},\bm{L}-\bm{\ell}}\\
0 & F^{\text{MV},EB}_{\bm{L}-\bm{\ell},\bm{\ell}} & 0
\end{pmatrix},
\eeq
with corrections going as $\mathcal{O}(L/\ell)$, where the matrix is in the basis $\{T,E,B\}$. The coefficients $(A,B,C)$ should be chosen to minimize Eq.~\eqref{eq:global_noise} in the large-lens limit:
\beq
\bal
\mathcal{N}^\kappa_{\bm{0}}
&=
2 (N^\kappa_{\bm{0}})^2
\int \ell\,d\ell \left[
\frac{1}{2} (\tilde{C}^{TT}_\ell)^2 A^2_{0,\ell} 
+
\frac{1}{2} (\tilde{C}^{EE}_\ell)^2 C^2_{0,\ell}
+ 
2 \big((\tilde{C}^{TE}_\ell)^2 + \tilde{C}^{TT}_\ell\tilde{C}^{EE}_\ell\big) B^2_{0,\ell} 
+ 
(\tilde{C}^{TE}_\ell)^2 A_{0,\ell}C_{0,\ell}
\right]\\
&+
2 (N^\kappa_{\bm{0}})^2
\int_{\bm{\ell}} 2 \tilde{C}^{EE}_\ell\tilde{C}^{BB}_\ell (F^{\text{MV},EB}_{\bm{\ell},-\bm{\ell}})^2,
\eal
\eeq
where
\beq
(N^\kappa_{\bm{0}})^{-1} 
=
\int_{\bm{\ell}}
2 f^{EB}_{\bm{\ell},-\bm{\ell}} F^{\text{MV},EB}_{\bm{\ell},-\bm{\ell}} 
+
\int \ell\,d\ell
\left[
\frac{1}{2}
\frac{d C^{EE}_\ell}{d\ln\ell}
C_{0,\ell}
+ 
2 
C^{TE}_\ell
\frac{d\ln \ell^2 C^{TE}_\ell}{d\ln\ell}
B_{0,\ell} 
+ 
\frac{1}{2}
\frac{d C^{TT}_\ell}{d\ln\ell}
A_{0,\ell}.
\right]
\eeq
By varying $A,B,C$ individually and setting $\delta \mathcal{N}^\kappa=0$ we find:
\beq
\bal
A_{L,\ell} &= 
\frac{
(\tilde{C}^{EE}_\ell)^2 \frac{dC^{TT}_\ell}{d\ln\ell}
-
(\tilde{C}^{TE}_\ell)^2 \frac{dC^{EE}_\ell}{d\ln\ell}
}{
2\left[(\tilde{C}^{TT}_\ell)^2 (\tilde{C}^{EE}_\ell)^2 - (\tilde{C}^{TE}_\ell)^4\right]
}
\quad
\quad
C_{L,\ell} = 
\frac{
(\tilde{C}^{TT}_\ell)^2 \frac{dC^{EE}_\ell}{d\ln\ell}
-
(\tilde{C}^{TE}_\ell)^2 \frac{dC^{TT}_\ell}{d\ln\ell}
}{
2\left[(\tilde{C}^{TT}_\ell)^2 (\tilde{C}^{EE}_\ell)^2 - (\tilde{C}^{TE}_\ell)^4\right]
}
\\
B_{L,\ell} &= 
\frac{ C^{TE}_\ell
\frac{d\ln\ell^2|C^{TE}_\ell|}{d\ln\ell}
}{
2\left[\tilde{C}^{TT}_\ell \tilde{C}^{EE}_\ell + (\tilde{C}^{TE}_\ell)^2\right].
}
\eal
\eeq
up to $\mathcal{O}(L/\ell)$ corrections. 

\begin{figure}[!h]
\centering
\includegraphics[width=0.4\linewidth]{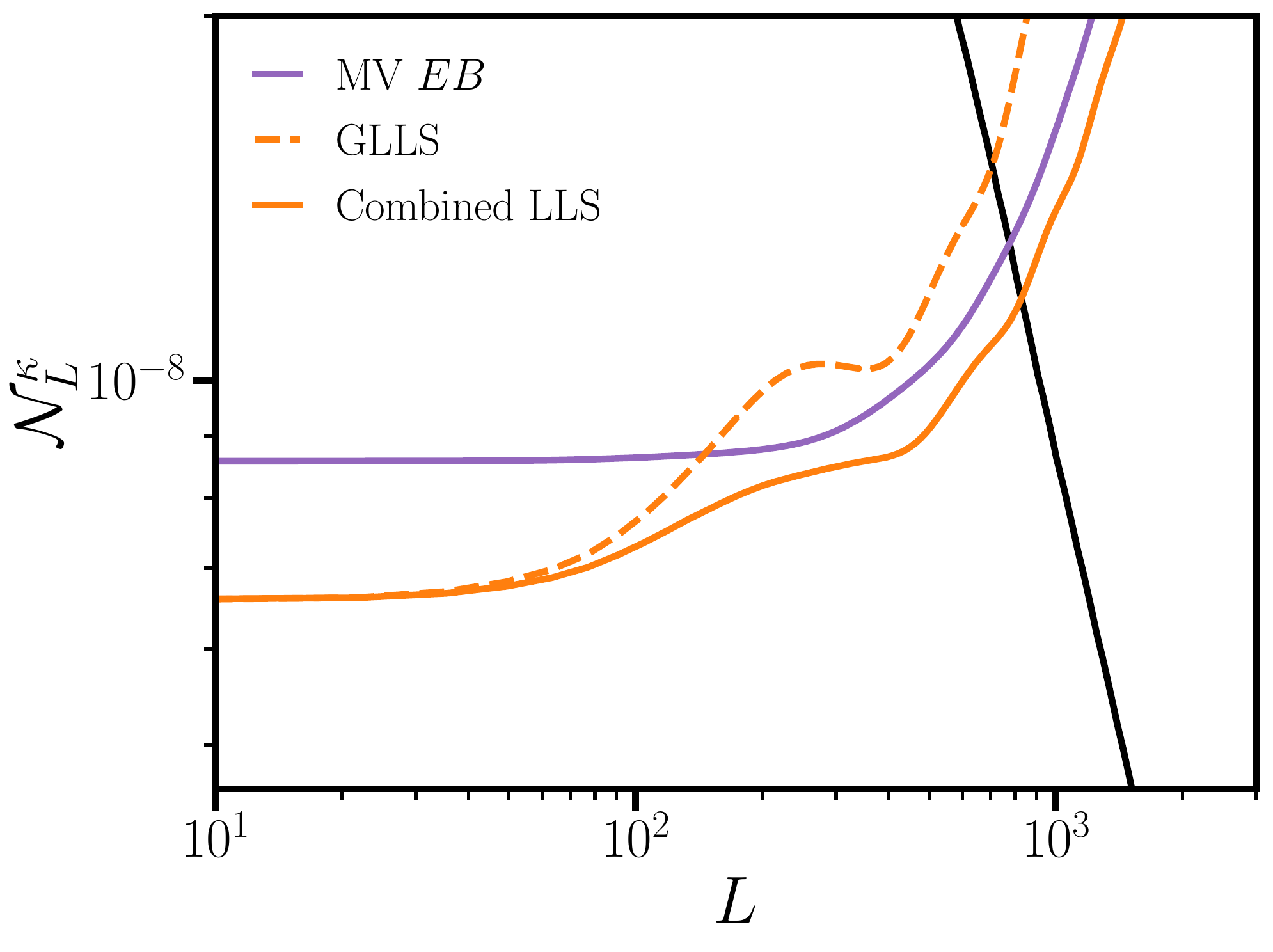}
\caption{Lensing reconstruction noise for the optimal linear combination of single-pair LLS estimators (solid orange) and for the Global LLS estimator (GLLS, dashed orange). For reference we show the reconstruction noise for the MV $EB$ estimator (purple) and the convergence power spectrum (black).
}
\label{fig:LLS_options}
\end{figure}

As shown in Fig.~\ref{fig:LLS_options}, the global LLS estimator (GLLS) negligibly improves upon the optimal linear combination of single-pair LLS estimators at large scales. At small scales where the $L/\ell\ll 1$ approximation breaks down the GLLS estimator is no longer guaranteed to be optimal. In fact we find that the optimal linear combination of LLS estimators outperforms (or is nearly identical to) the GLLS for all relevant scales.

\section{Bias hardening as constrained minimization}
\label{sec:optimal_bias_hardening}

Here we will derive an optimal bias hardened estimator using Lagrange multipliers. Let's suppose that we have a foreground $s^i$ that we wish to harden a generic quadratic estimator (Eq.~\eqref{eq:general_QE}) against, with the relevant linear responses given by $g^{ij,k}_{\bm{\ell},\bm{L}-\bm{\ell}}$. To construct an optimal bias hardened estimator, we choose our lensing weights $F^{ij}$ to minimize the noise of our lensing estimator under the constraint of having no response to the foreground. That is, we wish to minimize the Lagrangian: 
\beq
\label{eq:lagrangian_bh}
\mathcal{L}_{\bm{L}}[F^{ij}_{\bm{\ell},\bm{L}-\bm{\ell}},\lambda^i_{\bm{L}}]
=
\mathcal{N}^\kappa_{\bm{L}} + 2 \sum_i \lambda^i_{\bm{L}} \int_{\bm{\ell}} F^{jk}_{\bm{\ell},\bm{L}-\bm{\ell}} g^{jk,i}_{\bm{\ell},\bm{L}-\bm{\ell}}
\eeq
where $\lambda^i_{\bm{L}}$ are some Lagrange multipliers. We first vary the weights, which gives 
\beq
\delta \mathcal{L}_{\bm{L}} = 4 (N^\kappa_{\bm{L}})^2 \sum_{mn} \int_{\bm{\ell}}
\left[
\sum_{ij} F^{ij}_{\bm{\ell},\bm{L}-\bm{\ell}} \tilde{C}^{im}_\ell \tilde{C}^{jn}_{|\bm{L}-\bm{\ell}|}
- \frac{1}{2} \frac{\mathcal{N}^\kappa_{\bm{L}}}{N^\kappa_{\bm{L}}} f^{mn}_{\bm{\ell},\bm{L}-\bm{\ell}}
+ 
 \frac{1}{2}\sum_k \sigma^k_{\bm{L}} g^{mn,k}_{\bm{\ell},\bm{L}-\bm{\ell}}
\right] \delta F^{mn}_{\bm{\ell},\bm{L}-\bm{\ell}}
\eeq
where $\sigma^k_{\bm{L}} =  \lambda^k_{\bm{L}} / (N^\kappa_{\bm{L}})^2$ are a renormalized set of Lagrange multipliers. After setting $\delta\mathcal{L}_{\bm{L}} =0$ and fixing the normalization of the weights $(\mathcal{N}^\kappa_{\bm{L}} = N^\kappa_{\bm{L}})$, we find that the optimal zero-response weights satisfy
\beq
\bal
F^{ij}_{\bm{\ell},\bm{L}-\bm{\ell}} &=
\frac{1}{2}
\sum_{mn}
\left(\tilde{\bm{C}}^{-1}_\ell\right)^{im} \left(\tilde{\bm{C}}^{-1}_{|\bm{L}-\bm{\ell}|}\right)^{jn} 
\left[
f^{mn}_{\bm{\ell},\bm{L}-\bm{\ell}}
-
\sum_k \sigma^k_{\bm{L}}
g^{mn,k}_{\bm{\ell},\bm{L}-\bm{\ell}}
\right]
\\
&= F^{\text{GMV},ij}_{\bm{\ell},\bm{L}-\bm{\ell}} - \sum_k \sigma^k_{\bm{L}}
G^{ij,k}_{\bm{\ell},\bm{L}-\bm{\ell}}
\eal
\eeq
where $F^{\text{GMV},ij}$ are the weights for the GMV $\kappa$ estimator and $G^{ij,k}$ are the GMV weights for $s^k$. That is, the true minimum variance estimator with zero response to the foregrounds is some linear combination of the GMV estimators for $\kappa$ and $s^i$. The linear combination which nulls the response of $\hat{\kappa}$ to $s^i$ is unique, given by Eq.~\eqref{eq:bias_hardened}. 

This approach can in principle be extended to null not only the field-level bias (and hence the primary and tripectrum biases, see Section \ref{sec:bias}), but also the secondary bias. Consider the temperature-only case for simplicity. In this scenario the secondary bias is proportional to:
\beq
\text{Secondary}
\propto
\int_{\bm{\ell}\bm{\ell}'} 
F^{TT}_{\bm{\ell},\bm{L}-\bm{\ell}}
F^{TT}_{\bm{\ell}',-\bm{L}-\bm{\ell}'}
f^{TT}_{\bm{\ell},-\bm{L}-\bm{\ell}'}
g^{TT,T}_{\bm{\ell}',\bm{L}-\bm{\ell}}.
\eeq
We null the secondary by adding a second term to the Lagrangian, yielding:
\beq
F^{TT}_{\bm{\ell},\bm{L}-\bm{\ell}}
=
\frac{f^{TT}_{\bm{\ell},\bm{L}-\bm{\ell}}}{2 \tilde{C}^{TT}_\ell \tilde{C}^{TT}_{|\bm{L}-\bm{\ell}|}}
+
\sigma_{\bm{L}}
\frac{g^{TT,T}_{\bm{\ell},\bm{L}-\bm{\ell}}}{2 \tilde{C}^{TT}_\ell \tilde{C}^{TT}_{|\bm{L}-\bm{\ell}|}}
+
\frac{\lambda_{\bm{L}}}{2 \tilde{C}^{TT}_\ell \tilde{C}^{TT}_{|\bm{L}-\bm{\ell}|}}
\int_{\bm{\ell}'} 
F^{TT}_{\bm{\ell}',-\bm{L}-\bm{\ell}'}
f^{TT}_{\bm{\ell},-\bm{L}-\bm{\ell}'}
g^{TT,T}_{\bm{\ell}',\bm{L}-\bm{\ell}}
\eeq
where $\sigma$ and $\lambda$ are chosen to null the field-level and secondary biases respectively. In practice one would have to solve for these weights numerically, which could potentially be accomplished in an iterative fashion with the bias-hardened weights as an initial guess.

\section{Derivation of polarization-based LLS estimators}
\label{sec:LLS_details}
In a similar manner to \ref{sec:LLS} we expand our lensing weights into multipoles:
\beq
\bal
F^{XY}_{\bm{\ell},\bm{L}-\bm{\ell}} &= \sum_{m =0}^\infty A^{(m)}_{L,\ell} \cos(m \theta_{\bm{L},\bm{\ell}})
+ 
\sum_{m=1}^\infty
B^{(m)}_{L,\ell} \sin(m \theta_{\bm{L},\bm{\ell}}),
\eal
\eeq
The noise of the lensing estimator estimator in the large lens limit is then
\beq
\bal
\mathcal{N}_0 \propto \frac{\int \ell\,d\ell \tilde{C}^{XX}_\ell \tilde{C}^{YY}_\ell \left[2 \big(A^{(0)}_{0,\ell}\big)^2 + \sum_{m=1}^\infty  \big(A^{(m)}_{0,\ell}\big)^2 + \sum_{m=1}^\infty \big(B^{(m)}_{0,\ell}\big)^2 \right]}{
\left[
\sum_{m=0}^\infty A^{(m)}_{0,\ell} f^{XY,m}_{c,\ell} 
+
\sum_{m=1}^\infty B^{(m)}_{0,\ell} f^{XY,m}_{s,\ell} 
\right]^2
}.
\eal
\eeq
where we have neglected the cross-correlation $\tilde{C}^{XY}$, and we have defined
\beq
f^{XY,m}_{c,\ell} = \int \frac{d\theta_{\bm{L},\bm{\ell}}}{2\pi} \cos(m\theta_{\bm{L},\bm{\ell}}) f^{XY}_{\bm{\ell},-\bm{\ell}}
\quad
\quad
\quad
f^{XY,m}_{s,\ell} = \int \frac{d\theta_{\bm{L},\bm{\ell}}}{2\pi} \sin(m\theta_{\bm{L},\bm{\ell}}) f^{XY}_{\bm{\ell},-\bm{\ell}}.
\eeq
Varying $A^{(m)}$, $B^{(m)}$ individually and setting $\delta\mathcal{N}_0=0$ then gives
\beq
\label{eq:optimal_coefficients}
A^{(0)}_{L,\ell} \propto \frac{f^{XY,0}_{c,\ell}}{2\tilde{C}^{XX}_\ell \tilde{C}^{YY}_\ell } 
\quad
A^{(m)}_{L,\ell} \propto \frac{f^{XY,m}_{c,\ell}}{\tilde{C}^{XX}_\ell \tilde{C}^{YY}_\ell }  
\quad
B^{(m)}_{L,\ell} \propto \frac{f^{XY,m}_{s,\ell}}{\tilde{C}^{XX}_\ell \tilde{C}^{YY}_\ell } , 
\eeq
up to $\mathcal{O}(L/\ell)$, where the (common) proportionality constant is any function of $L$. That is, we can read off the optimal values of $A^{(m)},B^{(m)}$ directly from the multipole expansion of the linear response $f^{XY}$ in the large-lens limit. To suppress the response to foregrounds we will manually set the coefficients which are sensitive to foregrounds to zero. We do so below for each single-pair estimator, making frequent use of Eqs.~\eqref{eq:taylor_expanded_responses} and \eqref{eq:taylor_expanded_foreground_responses}.

$TE$: Note that $g^{TE,E}$ contains a $\cos(2\theta)$ term which is $\mathcal{O}(1)$, while $g^{TE,B}$ has a $\mathcal{O}(1)$ $\sin(2\theta)$ term. We therefore manually set $A^{(2)}=B^{(2)} = 0$ to suppress responses to $s^E$ and $s^B$. Both $g^{TE,E}$ and $g^{TE,E}$ only have a monopole at $\mathcal{O}((L/\ell)^2)$. Thus we can retain the monopole $A^{(0)}$, whose optimal value takes the form 
\beq
A^{(0)}_{L,\ell} = 
\frac{C^{TE}_\ell}{\tilde{C}^{TT}_\ell \tilde{C}^{EE}_\ell}\frac{d\ln \ell^2 |C^{TE}_\ell|}{d\ln\ell},
\eeq
where we have made us of Eqs.~\eqref{eq:optimal_coefficients} and \eqref{eq:taylor_expanded_responses}.

$TB$: Likewise $g^{TB,E}$ contains a $\mathcal{O}(1)$ $\sin(2\theta)$ term while $g^{TE,B}$ has a $\mathcal{O}(1)$ $\cos(2\theta)$ term. We again set $A^{(2)}=B^{(2)} = 0$ to suppress responses to $s^E$ and $s^B$. Both $g^{TE,E}$ and $g^{TE,E}$ only have a monopole at $\mathcal{O}((L/\ell)^2)$. Thus we can retain the monopole $A^{(0)}$, however, the response $f^{TB}$ only has a $\sin(2\theta)$ term to leading order, and so we set $A^{(0)}=0$. There is no LLS $TB$ estimator to leading order. 

$EE$: This case is identical to $TT$. We only retain the $\cos(2\theta)$ term.

$EB$: Note that $F^{\text{MV},EB} = f^{EB}/\tilde{C}^{EE} \tilde{C}^{BB}$ only has a $\sin(2\theta)$ term at $\mathcal{O}(1)$, while at $\mathcal{O}(L/\ell)$ it has a $\cos(\theta)$, $\sin(\theta)$ and $\sin(3\theta)$ term. Note also that $g^{EB,T}$ only has a $\sin(\theta)$ term at $\mathcal{O}(L/\ell)$. Thus the foreground response of the MV $EB$ estimator is already suppressed by $\mathcal{O}((L/\ell)^2)$.

\section{Worked example: explicitly hardening the $TE$ estimator}
\label{sec:explicit_te_psh}

From Table~\ref{tab:response} we see that the $TE$ has a non-zero linear response to $s^E$ and $s^B$ in addition to lensing. Thus to bias harden $\hat{\kappa}$ against $s^E$ and $s^B$ we must construct (optimal) estimators for $\kappa,s^E$ and $s^B$, which take the form:
\beq
\bal
\hat{\kappa}_{\bm{L}} &= N^\kappa_{\bm{L}}\int_{\bm{\ell}}
\frac{f^{TE}_{\bm{\ell},\bm{L}-\bm{\ell}}}{ \tilde{C}^{TT}_{\ell} \tilde{C}^{EE}_{|\bm{L}-\bm{\ell}|}} \tilde{T}_{\bm{\ell}} \tilde{E}_{\bm{L}-\bm{\ell}}
\quad
\text{where}
\quad
(N^\kappa_{\bm{L}})^{-1} = \int_{\bm{\ell}}
\frac{(f^{TE}_{\bm{\ell},\bm{L}-\bm{\ell}})^2}{ \tilde{C}^{TT}_{\ell} \tilde{C}^{EE}_{|\bm{L}-\bm{\ell}|}} 
\\
\hat{s}^E_{\bm{L}} &=  N^E_{\bm{L}}\int_{\bm{\ell}}
\frac{\cos(2\theta_{\bm{L},\bm{L}-\bm{\ell}})}{ \tilde{C}^{TT}_{\ell} \tilde{C}^{EE}_{|\bm{L}-\bm{\ell}|}} \tilde{T}_{\bm{\ell}} \tilde{E}_{\bm{L}-\bm{\ell}}
\quad
\text{where}
\quad
(N^E_{\bm{L}})^{-1} = \int_{\bm{\ell}}
\frac{\cos^2(2\theta_{\bm{L},\bm{L}-\bm{\ell}})}{ \tilde{C}^{TT}_{\ell} \tilde{C}^{EE}_{|\bm{L}-\bm{\ell}|}}
\\
\hat{s}^B_{\bm{L}} &= N^B_{\bm{L}}\int_{\bm{\ell}}
\frac{\sin(2\theta_{\bm{L}-\bm{\ell},\bm{L}})}{ \tilde{C}^{TT}_{\ell} \tilde{C}^{EE}_{|\bm{L}-\bm{\ell}|}} \tilde{T}_{\bm{\ell}} \tilde{E}_{\bm{L}-\bm{\ell}}
\quad
\text{where}
\quad
(N^B_{\bm{L}})^{-1} = \int_{\bm{\ell}}
\frac{\sin^2(2\theta_{\bm{L}-\bm{\ell},\bm{L}})}{ \tilde{C}^{TT}_{\ell} \tilde{C}^{EE}_{|\bm{L}-\bm{\ell}|}},
\eal
\eeq
where $\tilde{C}^{XX}_\ell$ is the total power spectrum, including the lensed field $\tilde{X}$ as well as instrumental noise and foregrounds. Here we have neglected the cross-correlation between $T$ and $E$ in the minimum variance weights to make the estimators FFT-able, and we have assumed that the foreground is composed of point sources ($u_\ell=1$). Note that we have neglected the normalization $\mathcal{B}$ of the foregrounds. This makes $\hat{s}^E$ an estimator of $ \mathcal{B}\, s^{E}$, and similarly for $\hat{s}^B$. As we show in Appendix \ref{sec:GPSH}, the bias hardened $TE$ lensing estimator is insensitive to $\mathcal{B}$, which we will also show below explicitly. 

The field-level biases to the minimum variance estimators above take the form:
\beq
\begin{pmatrix}
\langle\hat{\kappa}_{\bm{L}}\rangle'\\
\langle\hat{s}^E_{\bm{L}}\rangle'\\
\langle\hat{s}^B_{\bm{L}}\rangle'\\
\end{pmatrix}
=
\begin{pmatrix}
1 &  \mathcal{R}^{\kappa E}_{\bm{L}} & \mathcal{R}^{\kappa B}_{\bm{L}} \\
 \mathcal{R}^{E\kappa}_{\bm{L}} & 1 &  \mathcal{R}^{EB}_{\bm{L}} \\
\mathcal{R}^{B\kappa}_{\bm{L}} & \mathcal{R}^{BE}_{\bm{L}} & 1 \\
\end{pmatrix}
\begin{pmatrix}
\kappa_{\bm{L}}\\
\mathcal{B}\,s^E_{\bm{L}}\\
\mathcal{B}\,s^B_{\bm{L}}\\
\end{pmatrix},
\eeq
where we have defined six response functions:
\beq
\bal
\mathcal{R}^{\kappa E}_{\bm{L}} &=
N^\kappa_{\bm{\ell}}\int_{\bm{\ell}}
\frac{f^{TE}_{\bm{\ell},\bm{L}-\bm{\ell}}
\cos(2\theta_{\bm{L},\bm{L}-\bm{\ell}})
}{ \tilde{C}^{TT}_{\ell} \tilde{C}^{EE}_{|\bm{L}-\bm{\ell}|}}
\quad\quad\quad\quad\quad
\mathcal{R}^{E\kappa}_{\bm{L}} =
N^E_{\bm{\ell}}\int_{\bm{\ell}}
\frac{f^{TE}_{\bm{\ell},\bm{L}-\bm{\ell}}
\cos(2\theta_{\bm{L},\bm{L}-\bm{\ell}})
}{ \tilde{C}^{TT}_{\ell} \tilde{C}^{EE}_{|\bm{L}-\bm{\ell}|}}
\\
\mathcal{R}^{\kappa B}_{\bm{L}} &=
N^\kappa_{\bm{\ell}}
\int_{\bm{\ell}}
\frac{f^{TE}_{\bm{\ell},\bm{L}-\bm{\ell}}
\sin(2\theta_{\bm{L},\bm{L}-\bm{\ell}})
}{ \tilde{C}^{TT}_{\ell} \tilde{C}^{EE}_{|\bm{L}-\bm{\ell}|}}
\quad\quad\quad\quad\quad
\mathcal{R}^{B\kappa}_{\bm{L}} =
N^B_{\bm{\ell}}
\int_{\bm{\ell}}
\frac{f^{TE}_{\bm{\ell},\bm{L}-\bm{\ell}}
\sin(2\theta_{\bm{L},\bm{L}-\bm{\ell}})
}{ \tilde{C}^{TT}_{\ell} \tilde{C}^{EE}_{|\bm{L}-\bm{\ell}|}}
\\
\mathcal{R}^{E B}_{\bm{L}} &= 
N^E_{\bm{\ell}}
\int_{\bm{\ell}}
\frac{\cos(2\theta_{\bm{L},\bm{L}-\bm{\ell}})
\sin(2\theta_{\bm{L},\bm{L}-\bm{\ell}})
}{ \tilde{C}^{TT}_{\ell} \tilde{C}^{EE}_{|\bm{L}-\bm{\ell}|}}
\quad\quad
\mathcal{R}^{BE}_{\bm{L}} = 
N^B_{\bm{\ell}}
\int_{\bm{\ell}}
\frac{\cos(2\theta_{\bm{L},\bm{L}-\bm{\ell}})
\sin(2\theta_{\bm{L},\bm{L}-\bm{\ell}})
}{ \tilde{C}^{TT}_{\ell} \tilde{C}^{EE}_{|\bm{L}-\bm{\ell}|}}
.
\eal
\eeq
The bias hardened estimators are simply:
\beq
\begin{pmatrix}
\hat{\kappa}^\text{BH}_{\bm{L}}\\
\hat{s}^{\text{BH},E}_{\bm{L}}\\
\hat{s}^{\text{BH},B}_{\bm{L}}\\
\end{pmatrix}
=
\begin{pmatrix}
1 &  \mathcal{R}^{\kappa E}_{\bm{L}} & \mathcal{R}^{\kappa B}_{\bm{L}} \\
 \mathcal{R}^{E\kappa}_{\bm{L}} & 1 &  \mathcal{R}^{EB}_{\bm{L}} \\
\mathcal{R}^{B\kappa}_{\bm{L}} & \mathcal{R}^{BE}_{\bm{L}} & 1 \\
\end{pmatrix}^{-1}
\begin{pmatrix}
\hat{\kappa}_{\bm{L}}\\
\hat{s}^E_{\bm{L}}\\
\hat{s}^B_{\bm{L}}\\
\end{pmatrix}.
\eeq

\section{Expected biases to Simons Observatory}
\label{sec:SO_biases}
\begin{figure}[!h]
\centering
\includegraphics[width=0.5\linewidth]{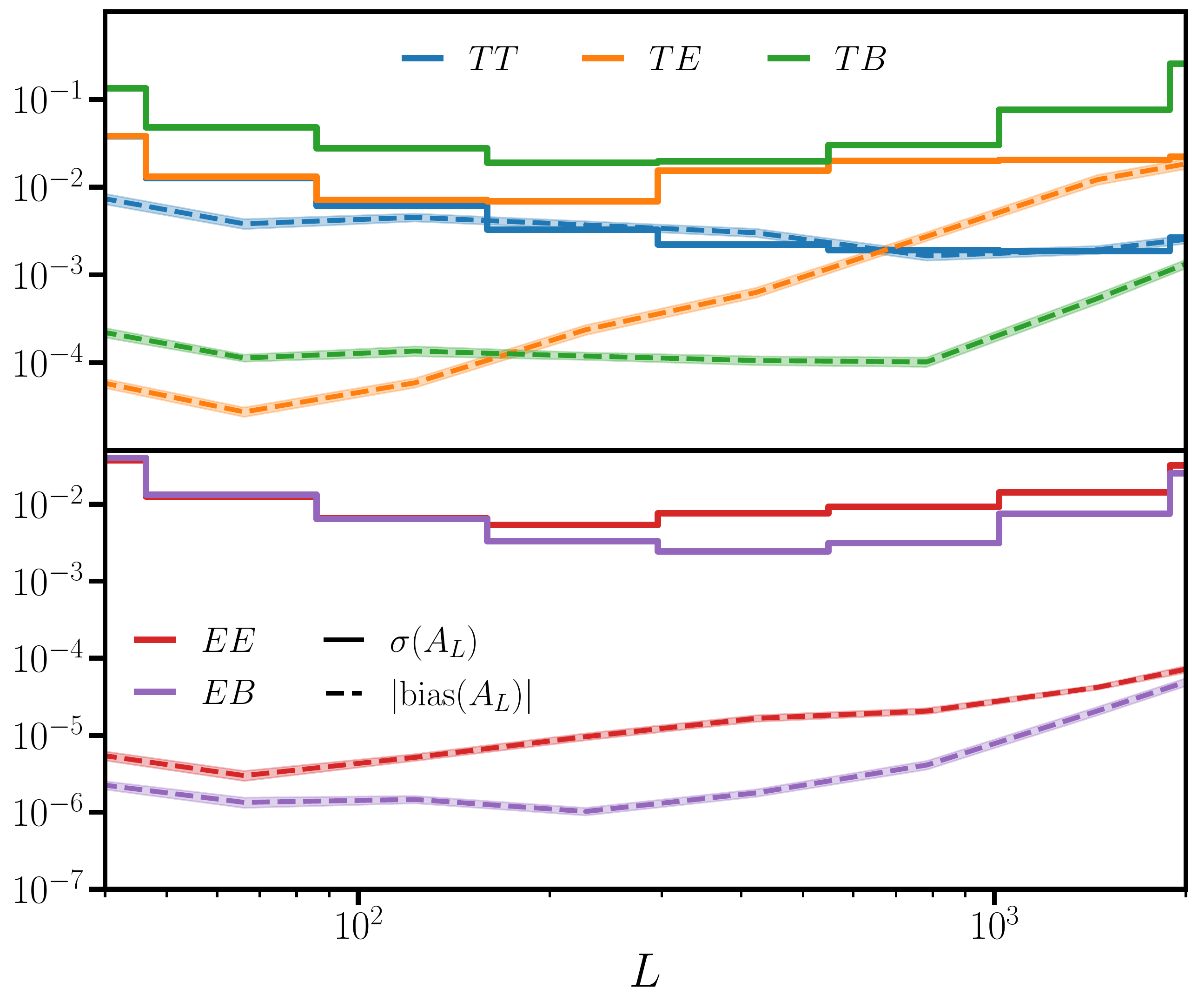}
\caption{The same as the top two panels of Fig.~\ref{fig:bias_and_noise}, but with Simons Observatory-like sensitivity instead of CMB-S4.
}
\label{fig:bias_and_noise_SO}
\end{figure}

In Fig.~\ref{fig:bias_and_noise_SO} we show the expected sensitivity to the lensing amplitude, and the bias from radio point sources, for  Simons Observatory-like sensitivity. The assumptions regarding the experimental setup and simulations are identical to those in the main text, but with $\Delta_T =6$ $\mu \text{K}$-arcmin instead of $\Delta_T =1$ $\mu \text{K}$-arcmin. We see that for polarization-based reconstruction the point source-induced bias is negligible.

\end{document}